\documentclass[longauth]{aa}  
\usepackage{graphicx}
\usepackage[varg]{txfonts}
\usepackage{gensymb}
\usepackage{subcaption}
\usepackage{hyperref}
\usepackage{placeins}   

\begin{document} 

   \title{Complex Organic Molecules in Protostars with ALMA Spectral Surveys (COMPASS)}
    \subtitle{III. CH$_3$OH isotopic fractionation in the low-mass protostar BHR71-IRS1}
    \titlerunning{CH$_3$OH isotopic fractionation in the low-mass protostar BHR71-IRS1}

   \author{A. Coutens \inst{1}, F. Cruz-S\'aenz de Miera \inst{1}, J. K. J{\o}rgensen\inst{2}, 
   M. N. Drozdovskaya\inst{3,4}, J.-E. Lee\inst{5}, A. L. Plunkett\inst{6}, D. Harsono\inst{7}, S. Maret\inst{8}, 
   A. Belloche\inst{9},  J. Ferrer Asensio\inst{10}, M. L. R. van't Hoff\inst{11}, J.-H. Jeong\inst{5}, \'A.  K\'osp\'al\inst{12,13,14}, B. A. McGuire\inst{15,6}, P. Nazari\inst{16},  S. Spezzano\inst{17},  Z. Telkamp\inst{18}, Y.-L. Yang\inst{10}, H.-S. Yun\inst{5,19}, S. Zeng\inst{10}, 
          A. Andreu\inst{1}, J. Bergner\inst{20}, C.-H. Kim\inst{5}, N. F. W. Ligterink\inst{21}, Y. Lin\inst{17}, S.-Y. Liu\inst{22}, M. L{\"u}tzen\inst{2}, P. Marchand\inst{1}, M. Rao\inst{18}}
   \authorrunning{A. Coutens et al.}

   \institute{Institut de Recherche en Astrophysique et Plan\'etologie, Universit\'e de Toulouse, CNRS, CNES, 9 av. du Colonel Roche, 31028 Toulouse Cedex 4, France, \email{audrey.coutens@utoulouse.fr}
         \and Niels Bohr Institute, University of Copenhagen, Jagtvej 155A, DK-2200 Copenhagen N, Denmark
         \and Physikalisch-Meteorologisches Observatorium Davos und Weltstrahlungszentrum (PMOD/WRC), Dorfstrasse 33, 7260 Davos Dorf, Switzerland      
         \and Department of Chemistry, Biochemistry and Pharmaceutical Sciences (DCBP), Universit\"at Bern, Freiestrasse 3, 3012 Bern,
Switzerland
         \and Department of Physics and Astronomy, Seoul National University, 1 Gwanak-ro, Gwanak-gu, Seoul 08826, Korea        
         \and National Radio Astronomy Observatory, 520 Edgemont Road, Charlottesville, VA 22903, USA
         \and Institute of Astronomy, Department of Physics, National Tsinghua University, No. 101, Section 2, Kuang-Fu Road, Hsinchu 30013, Taiwan, R.O.C
         \and Univ. Grenoble Alpes, CNRS, IPAG, 38000 Grenoble, France
         \and Max-Planck-Institut f{\"u}r Radioastronomie, Auf dem H{\"u}gel 69, 53121 Bonn, Germany     
         \and RIKEN, Star and Planet Formation Laboratory, RIKEN Pioneering Research Institute, Wako-shi, Saitama, 351-0198, Japan    
       	\and Purdue University, 525 Northwestern Ave, West Lafayette, IN 47907 USA
         \and Konkoly Observatory, HUN-REN Research Centre for Astronomy and Earth Sciences, MTA Centre of Excellence, Konkoly-Thege Mikl\'os \'ut 15-17, 1121 Budapest, Hungary 
         \and Institute of Physics and Astronomy, ELTE E\"otv\"os Lor\'and University, P\'azm\'any P\'eter s\'et\'any 1/A, 1117 Budapest, Hungary 
         \and Max-Planck-Institut f\"ur Astronomie, K\"onigstuhl 17, 69117 Heidelberg, Germany   
         \and Department of Chemistry, Massachusetts Institute of Technology, 77 Massachusetts Ave, Cambridge, MA 02139 USA
         \and European Southern Observatory, European Southern Observatory, Karl-Schwarzschild-Strasse 2, 85748 Garching, Germany  
         \and Max-Planck-Institut f{\"u}r extraterrestrische Physik, Giessenbachstrasse 1, 85748 Garching bei M{\"u}nchen, Germany             
         \and Department of Astronomy, University of Virginia, 530 McCormick Road, Charlottesville, VA 22904, US
         \and Korea Astronomy and Space Science Institute, 776 Daedeok-daero, Yuseong-gu, Daejeon 34055, Republic of Korea
         \and UC Berkeley, Department of Chemistry, Berkeley, CA 94720, USA
         \and Faculty of Aerospace Engineering, Delft University of Technology, Delft, Kluyverweg 1, 2629 HS Delft, The Netherlands
         \and Institute of Astronomy and Astrophysics, Academia Sinica, 11F of AS/NTU Astronomy-Mathematics Building, No.1, Sec. 4, Roosevelt Rd, Taipei 10617, Taiwan, R.O.C.
             }

   \date{Received xxx; accepted xxx}
 
  \abstract 
{Methanol is a complex organic molecule detected in both ice and gas toward star-forming regions. Due to its high abundance, its rarer isotopologues, including the deuterated ones, are also commonly detected. 
Measurements of deuteration can provide key constraints on the formation and evolution of molecules during the formation of stars and planets. It is not clear how the methanol deuteration varies between star-forming regions. Thanks to several recent spectroscopic studies, it is now possible to precisely derive the column densities of five deuterated isotopologues of methanol. }
{Our aim is to determine the column density and deuterium fractionation of methanol in the low-mass protostar BHR71-IRS1.}
{We analyzed data from a large spectral survey of BHR71-IRS1 carried out as part of the ALMA Large Program Complex Organic Molecules in Protostars with ALMA Spectral Surveys (COMPASS).  
We used the CASSIS software to identify the lines of the methanol isotopologues and constrain their column densities under the assumption of local thermodynamic equilibrium.}
{We detected CH$_3$OH, $^{13}$CH$_3$OH, CH$_3$$^{18}$OH, CH$_3$OD, CH$_2$DOH, CHD$_2$OH, and CD$_3$OH at the continuum peak position. In addition, CH$_3$$^{17}$OH is tentatively detected, while CD$_3$OD is not detected, although a few faint lines ($\lesssim$3$\sigma$) could be present toward a position redshifted from the continuum peak. The column densities of the main isotopologue obtained from $^{13}$CH$_3$OH and CH$_3$$^{18}$OH are consistent, assuming nominal $^{12}$C/$^{13}$C and $^{16}$O/$^{18}$O ratios of 68 and 557, respectively. The CH$_2$DOH/CH$_3$OD column density ratio is equal to 7$\pm$2. 
The statistically corrected D/H ratio is higher for CH$_2$DOH (1.9$\pm$0.6\,\%) than for CH$_3$OD (0.8$\pm$0.2\,\%). The level of deuteration after statistical correction increases with the number of D atoms in the methyl group, reaching 23$\pm$5\,\% for CD$_3$OH/CH$_3$OH. }
{These trends are similar to other low-mass protostars such as IRAS~16293--2422 and HOPS 373SW, but seem to differ from the eruptive young star V883-Ori. Similar studies with comparable datasets will be carried out for the other COMPASS sources to determine if factors such as environmental conditions and evolutionary states impact the deuteration of methanol. }


   \maketitle

\section{Introduction}

Methanol (CH$_3$OH) is the most abundant complex organic molecule (COM; i.e., carbon-containing molecule composed of six or more atoms) in the interstellar medium \citep{Herbst2009}. 
It is commonly found in the gas phase in a large variety of regions related to star and planet formation, such as protostars, shocks, prestellar cores, protoplanetary disks, and comets \citep[e.g.,][]{Menten1988,Goldsmith1999,Biver2002,Jorgensen2005,Maret2005,Walsh2016,Holdship2019,Scibelli2020,vanGelder2022}. 
Gas-phase formation routes are not sufficient to explain the high abundances of methanol in protostars \citep[e.g.,][]{Geppert2006}, which suggests an efficient formation on grains followed by thermal desorption in the warm inner regions of protostars, also referred to as hot corinos in the case of low-mass protostars \citep[e.g.,][]{Watanabe2002,Schoier2002,Maret2005,Fuchs2009}. 
This scenario is consistent with the detection of methanol ices in cold regions of molecular clouds \citep[e.g.,][]{Skinner1992,Boogert2008,Yang2022,McClure2023}. 

Due to its high abundance, methanol is the COM for which we can most easily derive its level of deuteration. The deuteration of molecules is highly sensitive to the physical conditions under which they form, making it a valuable tracer of the chemical evolution of star-forming regions from cold prestellar cores to warmer protostellar environments \citep{Caselli2012,Jorgensen2020}. In solar-type star-forming regions, the deuterium fractionation pathways are very efficient, and they led to the discoveries of not only the monodeuterated isotopologues (CH$_2$DOH, CH$_3$OD) but also the doubly and triply deuterated isotopologues (CHD$_2$OH, CD$_3$OH) with single-dish telescopes as early as the 2000s \citep{Parise2002,Parise2004,Parise2006}. Many questions arose regarding the deuterium fractionation processes of methanol. For example, the CH$_2$DOH/CH$_3$OD ratio seems to significantly deviate from the statistically expected 3:1 ratio, with values either higher or lower, depending on the luminosity of the source \citep[e.g.,][]{Ratajczak2011}, suggesting that the deuteration of the methyl and the hydroxyl groups could involve different processes \citep[e.g.,][]{Kulterer2022}. 

Within the last decade, interferometric studies have been performed to study the deuterium fractionation of methanol at small spatial scales. With the Protostellar Interferometric Line Survey (PILS; \citealt{Jorgensen2016}), a large spectral survey of the Class 0 protostellar system IRAS~16293--2422 (hereafter IRAS16293) with the Atacama Large Millimeter/Submillimeter Array (ALMA), a large number of transitions of various isotopologues of methanol were detected, which made it possible to precisely constrain the deuterium fractionation of methanol in this source. Indeed, the survey was sensitive enough to detect the CH$_3$$^{18}$OH isotopologue, which was used to determine the CH$_3$OH column density \citep{Jorgensen2016}, given the opacity issues with the main isotopologue \citep[e.g.,][]{Cruz2025}. 
Among the deuterated isotopologues detected toward component B of this source, there are CH$_2$DOH \citep{Jorgensen2016}, CH$_3$OD \citep{Ilyushin2024}, CHD$_2$OH \citep{Drozdovskaya2022}, and CD$_3$OH \citep{Ilyushin2022}. CD$_3$OD has also been tentatively detected \citep{Ilyushin2023}. The D/H ratio of the monodeuterated forms of methanol is higher for the methyl group ($\sim$2.4\%) than for the hydroxyl group ($\sim$0.32\%), as suggested in other studies \citep{Taquet2019}. The doubly and triply deuterated forms (CHD$_2$OH and CD$_3$OH) show even higher D/H ratios with respect to CH$_3$OH ($\sim$8\% and $\sim$15\%, respectively) than CH$_2$DOH.
 The CH$_2$DOH/CH$_3$OH and CHD$_2$OH/CH$_3$OH abundance ratios derived toward component A of IRAS16293 by \citet{Manigand2020} are consistent with those of component B. 
It is thus clear that the deuteration efficiency increases with the number of D atoms in IRAS16293. 
While studies of other sources are available for specific isotopologues, especially CH$_2$DOH \citep[e.g.,][]{Taquet2019,Hsu2022}, there is no other source with such a complete study of the D/H ratios from monodeuterated to triply deuterated methanol. Only HOPS 373SW shows detections of CH$_2$DOH, CH$_3$OD, CHD$_2$OH, and CD$_3$OH \citep{Lee2023}, but the number of lines (from 1 for CH$_3$OD to 14 for CH$_2$DOH) used to constrain the column densities is considerably lower than for IRAS16293 due to the narrower spectral coverage. 

With the ALMA Large Program called Complex Organic Molecules in Protostars with ALMA Spectral Surveys (COMPASS; \citealt{Jorgensen2026}), we carried out full spectral surveys of eleven protostellar regions from 279 to 312 GHz. The aim was to characterize the chemical composition of these regions and evaluate how the environment and evolution contribute to the observed differences. Here, we present an analysis of the D/H ratios for all the CH$_3$OH deuterated isotopologues available in the public spectroscopic databases (CH$_2$DOH, CH$_3$OD, CHD$_2$OH, CD$_3$OH, CD$_3$OD) toward BHR71-IRS1 (hereafter BHR71IRS1), which is the first source that was fully observed in this Large Program.  

BHR71IRS1 is a Class 0 protostar located in a relatively isolated region near the Southern Coalsack at a distance of about 176 pc \citep{Tobin2018,Voirin2018}. It has a binary component called BHR71-IRS2, which is $\sim$16$\arcsec$ away toward the west. Clear outflows traced by SiO and CO have been detected in both sources \citep[e.g.,][]{Parise2006,Gavino2024}. In BHR71IRS1, the direction of the outflow is close to a north-south alignment, with blueshifted emission toward the south and redshifted emission toward the north. 
A few transitions of CH$_3$OH, $^{13}$CH$_3$OH, CH$_2$DOH, and CH$_3$OD have already been identified in BHR71IRS1 with ALMA by \citet{Yang2020}. Singly deuterated water (HDO) has also been detected in this source \citep{Jensen2019}. This source shows an HDO/H$_2$O a factor of two higher than in IRAS16293 and a factor of two to three higher than three protostars belonging to the NGC1333 cloud. The higher water deuteration ratio of BHR71IRS1 could be explained by its more isolated environment compared to the other sources \citep{Jensen2021a}. Here, we investigate if this is also the case for methanol.

The paper is organized as follows. In Sect. \ref{sect_obs}, the COMPASS observations are described. In Sect. \ref{sect_analysis}, we present the methodology for the spectral analysis (with supplementary information in Appendix \ref{app_free_parameters}) and derive column densities for the CH$_3$OH isotopologues.  Section \ref{sect_discussion} is dedicated to the comparison with other sources and molecules from the literature. We conclude in Sect.~\ref{sect_conclusion}.

\section{Observations}
\label{sect_obs}

The observations were carried out with the ALMA 12m array in the framework of the COMPASS Large Program (2022.1.00316.L; \citealt{Jorgensen2026}).\footnote{\url{https://erda.ku.dk/vgrid/COMPASS/}} Large spectral surveys between 279 and 312 GHz were conducted toward eleven sources with a spectral resolution of 0.488 MHz, i.e., $\sim$0.5 km\,s$^{-1}$. BHR71IRS1 was observed in January-February 2023. The COMPASS data were calibrated with CASA, then imaged with Imager\footnote{\url{https://imager.oasu.u-bordeaux.fr/}}  for its flexibility and speed \citep{Plunkett2026}. Comparisons with the delivered ALMA standard products and a parallel reduction with CASA for this source show consistent results \citep{Plunkett2026}. The synthesized beam sizes for the nine settings used to produce the whole spectral survey are quite similar, ranging from 0$\farcs$42\,$\times$\,0$\farcs$33 to 0$\farcs$51\,$\times$\,0$\farcs$38. The rms varies from 1.9 to 2.9 mJy\,beam$^{-1}$ per channel with an average value of 2.2 mJy\,beam$^{-1}$ per channel. Spectral overlaps between the settings confirm that the calibration is consistent within 10\%. Continuum images were produced for each spectral window by identifying line-free channels in the $uv$-plane. The continuum values were later included in the radiative transfer model calculations (see Sect. \ref{sect_LTE}). At the continuum peak position, they vary between 81 and 101 mJy\,beam$^{-1}$ (36--52 K), in connection with the different beam sizes. 
We found that the standard continuum identification tools in, for example, CASA and IMAGER give sufficiently good estimates for this work; however, a small over-subtraction takes place in the line confused regions at the level of two to three times the channel rms. Therefore, we applied a pixel-by-pixel continuum subtraction method similar to the one presented in \citet{Jorgensen2016} to optimize the extracted spectra.  
The spectra were then extracted with a Python script that takes into account primary beam correction. More information about the data reduction can be found in \citet{Plunkett2026}. The COMPASS data reduction pipeline is archived on Zenodo \citep{Maret2025}.

\section{Results}
\label{sect_analysis}

\subsection{Extraction of the spectra}

\begin{figure}[t!]
\begin{center}
\includegraphics[width=7.5cm]{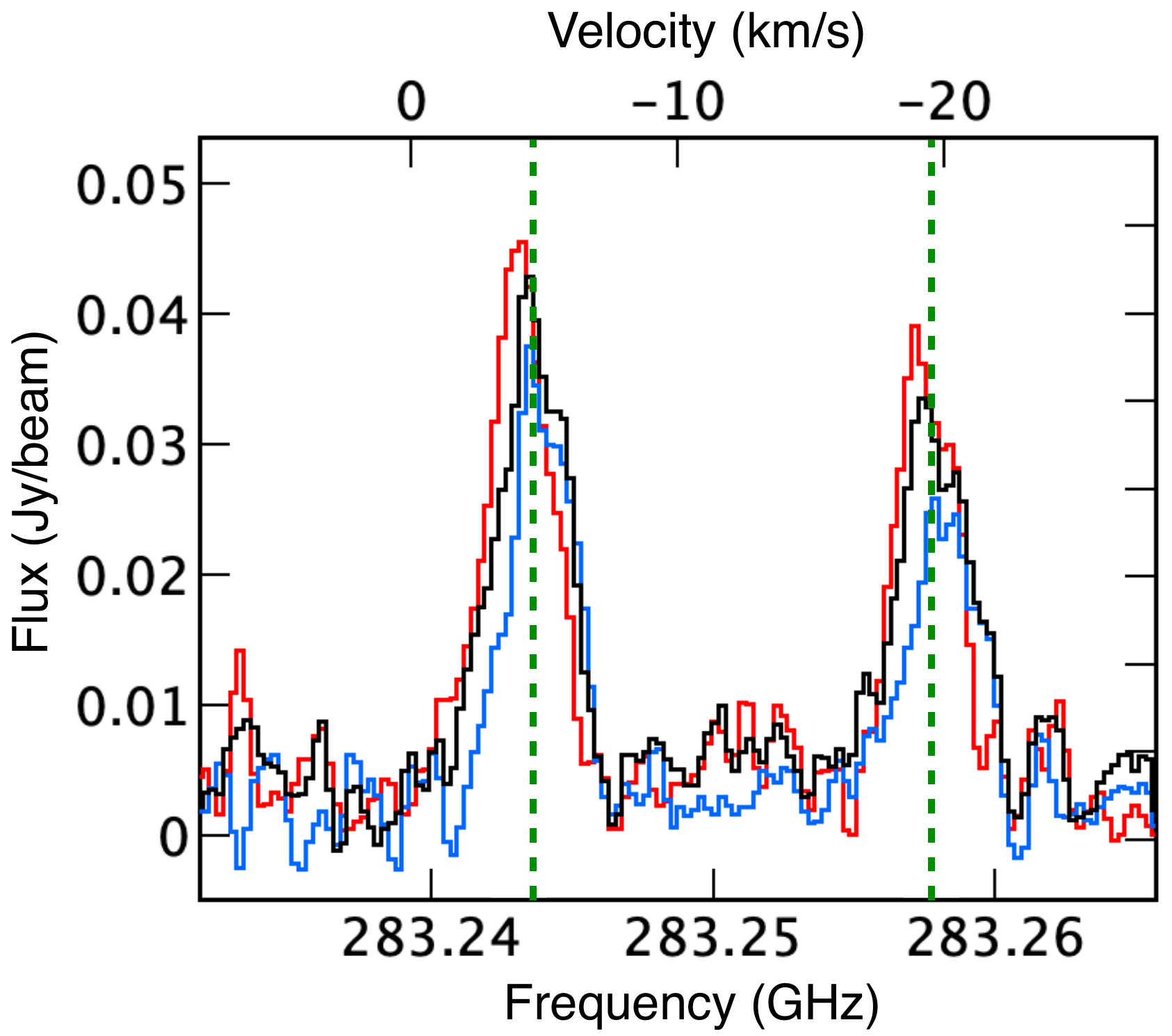}
\caption{Comparison of the profiles of two $^{13}$CH$_3$OH lines (283.2435 GHz with $E_{\rm up}$ = 69 K and $A_{\rm ij}$ = 1\,$\times$\,10$^{-4}$ s$^{-1}$ and 283.2576 GHz with $E_{\rm up}$ = 72 K and $A_{\rm ij}$ = 9\,$\times$\,10$^{-5}$ s$^{-1}$) at the continuum peak position (black), the blueshifted position (blue), and the redshifted position (red) indicated in Figs. \ref{fig_maps_moment0} and \ref{fig_maps_moment1}. The dashed green lines indicate the frequencies of the $^{13}$CH$_3$OH transitions at the systemic velocity of the source ($-$4.45 km\,s$^{-1}$, \citealt{Bourke1997}). The velocity axis indicated on top was calculated using the frequency of the first transition. }
\label{fig_comp_profiles}
\end{center}
\end{figure}

To identify the optimal places to extract spectra and derive column densities, we explored how the spectra change at different positions in the cubes using the Aladin\footnote{\url{https://aladin.cds.unistra.fr/AladinDesktop/}} software combined with the CASSIS$\footnote{CASSIS has been developed by IRAP-UPS/CNRS \citep{Vastel2015}. \url{https://cassis.irap.omp.eu}}$ plugin \citep{Glorian2024}. 
The line profiles of complex organic molecules seen at the dust continuum peak position seem to result from a combination of a blueshifted component and a redshifted one, as shown for two $^{13}$CH$_3$OH transitions in Fig. \ref{fig_comp_profiles}.
We consequently extracted the spectra at three different pixels. The first one corresponds to the dust continuum peak position ($\alpha$(ICRS)=12$^{\rm h}$01$^{\rm m}$36$\fs$476, $\delta$(ICRS)=$-$65$\degree$08$\arcmin$49.37$\arcsec$) determined by \citet{Gavino2024} using high angular resolution (0$\farcs$073\,$\times$\,0$\farcs$053) continuum images at 1.3 mm as part of the eDisk program \citep{Ohashi2023}. The two other pixels correspond to positions where the emission is either blueshifted or redshifted. They were chosen by eye to optimize the signal-to-noise ratio for a maximum number of lines whatever the molecule. The coordinates of the redshifted position are $\alpha$(ICRS)=12$^{\rm h}$01$^{\rm m}$36$\fs$460, $\delta$(ICRS)=$-$65$\degree$08$\arcmin$49.25$\arcsec$, while the coordinates of the blueshifted position are $\alpha$(ICRS)=12$^{\rm h}$01$^{\rm m}$36$.^{\rm s}$490, $\delta$(ICRS)=$-$65$\degree$08$\arcmin$49.49$\arcsec$. The blueshifted and redshifted positions are separated by 0$\farcs$31 from each other and by $\sim$0$\farcs$15 from the continuum peak position. 
Even if the positions are separated by less than a beam width, meaningful differences can still arise. However, these measurements are subject to beam dilution and blending effects, and care must be taken when interpreting small-scale variations, as they may not correspond to fully resolved, physically distinct regions. The three positions can be seen on the maps presented in Figs. \ref{fig_maps_moment0}--\ref{fig_maps_moment1}. Regarding the redshifted position, the emission lines (see Fig. \ref{fig_comp_profiles}) appear to be slightly brighter and narrower than toward the continuum peak position. This could potentially help distinguish partially blended transitions, but it should also be noted that the line profiles are not as Gaussian as toward the continuum peak position.

\subsection{Line identification}
\label{sect_line_id}

For the line identification, we used the CASSIS software. The spectroscopic data for the different methanol isotopologues come from the Cologne Database for Molecular Spectroscopy (CDMS; \citealt{Muller2001,Muller2005}), except for CH$_2$DOH, which is only available in the Jet Propulsion Laboratory (JPL) database \citep{Pickett1998}. More information about the spectroscopic studies of these species is available in Appendix \ref{app_spectro}. Since the spectral survey of BHR71IRS1 is line-rich, we carried out a preliminary identification of the lines in the full survey using local thermodynamic equilibrium (LTE) models. In this way, we obtained a first inventory of the species present in this source. 
In a second step, we checked if the lines assigned to the methanol isotopologues suffer from contamination with transitions of other molecules. For the analysis, we only considered the lines that were found to be unblended. The lines that are obviously blended with unidentified species (very broad profiles and/or significantly velocity-shifted peaks) were also excluded.

CH$_3$OH, $^{13}$CH$_3$OH, CH$_3$$^{18}$OH, CH$_2$DOH, CH$_3$OD, CHD$_2$OH, and CD$_3$OH are clearly detected toward BHR71IRS1 (see Figs. \ref{fig_model_13CH3OH}-\ref{fig_model_CD3OH}). For CH$_3$$^{17}$OH, we systematically see emission features at the frequencies of the brightest transitions of this molecule. However, most of them are partially blended with other molecules or relatively noisy, which makes its analysis more complicated than for the other isotopologues. Therefore, we consider it a tentative detection (see Fig. \ref{fig_model_CH317OH}). On the continuum peak and blueshifted position, CD$_3$OD is not detected, but, on the redshifted position, we see a better agreement between the models and the observations, suggesting that there could be CD$_3$OD lines slightly below 3$\sigma$ at this position (see Fig. \ref{fig_model_CD3OD}). We derived column density upper limits for this species.

\subsection{Maps}
\label{sect_maps}

\begin{figure*}[h!]
\begin{center}
\includegraphics[width=17cm]{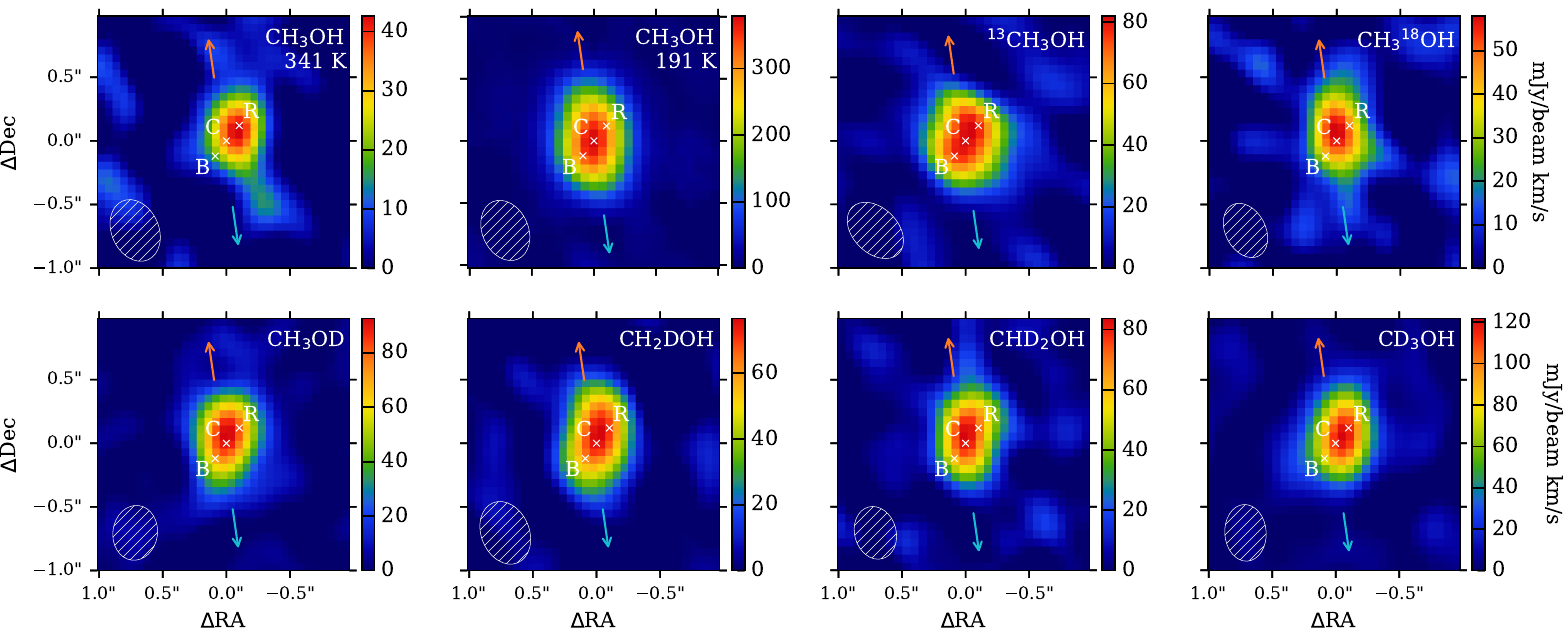}
\caption{
Moment 0 maps of the transitions of CH$_3$OH at 282213.3 MHz ($E_{\rm up}$ = 341\,K, $A_{\rm ij}$ = 9\,$\times$\,10$^{-7}$\,s$^{-1}$) and 281000.1 MHz ($E_{\rm up}$ = 191\,K, $A_{\rm ij}$ = 4\,$\times$\,10$^{-5}$\,s$^{-1}$), $^{13}$CH$_3$OH at 283207.3 MHz (two transitions with $E_{\rm up}$ =\,95 and 110 K, $A_{\rm ij}$ = 7 \,$\times$\,10$^{-5}$\,s$^{-1}$), CH$_3^{18}$OH at 309979.3 MHz ($E_{\rm up}$ = 77\,K, $A_{\rm ij}$ = 3\,$\times$\,10$^{-4}$\,s$^{-1}$), CH$_3$OD at 291412.8 MHz ($E_{\rm up}$ = 161\,K, $A_{\rm ij}$ = 1\,$\times$\,10$^{-4}$\,s$^{-1}$), CH$_2$DOH at 279335.7 MHz ($E_{\rm up}$ = 132\,K, $A_{\rm ij}$ = 3\,$\times$\,10$^{-5}$\,s$^{-1}$), CHD$_2$OH at 296460.6 MHz ($E_{\rm up}$ = 108\,K,  $A_{\rm ij}$ = 7\,$\times$\,10$^{-5}$\,s$^{-1}$), and CD$_3$OH at 305957.8 MHz ($E_{\rm up}$ = 163\,K, $A_{\rm ij}$ = 2\,$\times$\,10$^{-4}$\,s$^{-1}$). The integration was done between -9 and 0 km\,s$^{-1}$. The continuum peak (C) and the blueshifted (B) and the redshifted (R) positions are indicated with crosses. The directions of the blueshifted and redshifted outflows are indicated with cyan and orange arrows, respectively. The 0,0 position is the continuum peak position. }
\label{fig_maps_moment0} 
\end{center}
\end{figure*}

\begin{figure*}[h!]
\begin{center}
\includegraphics[width=17cm]{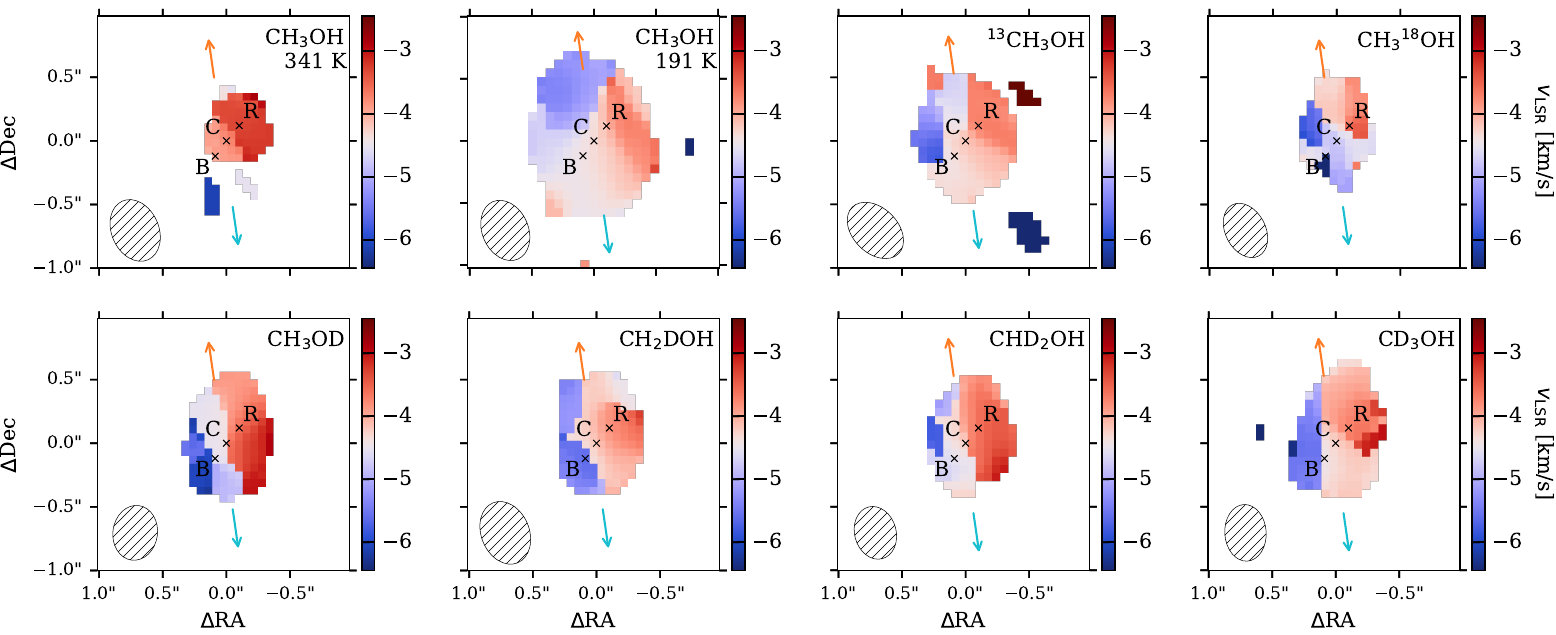}
\caption{Intensity weighted velocity (moment 1) maps of the methanol transitions shown in Fig. \ref{fig_maps_moment0}, created with a 4$\sigma$ cutoff. The systemic velocity is $-$4.45 km\,s$^{-1}$. The marked positions and arrows are the same as in Fig. \ref{fig_maps_moment0}. }
\label{fig_maps_moment1}
\end{center}
\end{figure*}

Moment 0 maps of transitions of the different isotopologues are presented in Fig. \ref{fig_maps_moment0}. The transitions were selected among the ones used in the $\chi^2$ minimization (see Sect. \ref{sect_LTE} and \href{url}{https://doi.org/10.5281/zenodo.19595973}) that are well fitted by the best-fit models, so they should not be blended. For a better comparison, we chose those with a similar $E_{\rm up}$  between 100 and 200\,K when possible. For CH$_3$$^{18}$OH, the highest $E_{\rm up}$ is 77 K, so we chose this line. For CH$_3$OH, we used the well fitted line with the lowest $E_{\rm up}$,  341\,K. We included this line, as well as a highly optically thick transition with an $E_{\rm up}$ of 191 K for comparison. The emission of the lines of the different isotopologues is compact and not spatially resolved. Their spatial distribution is very similar for all isotopologues. Only the CH$_3$OH line with $E_{\rm up}$ = 341 K looks slightly more compact, which is probably due to the higher $E_{\rm up}$ value and low $A_{\rm ij}$ (9\,$\times$\,10$^{-7}$ s$^{-1}$) of the transition. This more compact emission is also seen for the other transitions of CH$_3$OH with very high $E_{\rm up}$. A high-spatial resolution map from the e-Disk project confirms that the methanol emission is compact and concentrated in the inner region \citep{Sharma2025}. However, it should be noted that not all the methanol lines are confined to the warm inner region of the source. A more extended and complex spatial distribution with emission along the outflows is seen for bright CH$_3$OH lines with low $E_{\rm up}$ ($\lesssim$100\,K, \citealt{Yun2026}). These lines are not studied here due to their extremely high optical depths toward the selected positions. 

The moment 1 maps in relation to the position of the outflow axis \citep{Gavino2024} are shown in Fig. \ref{fig_maps_moment1}.
They reveal a velocity gradient approximately perpendicular to the outflow axis. It is likely associated with the kinematics of the large scale infalling and rotating envelope; a rotationally supported (Keplerian) disk has not been unambiguously identified yet toward this source, not even from the higher resolution eDisk data \citep{Gavino2024}.
The transition of CH$_3$OH with $E_{\rm up}$ = 341\,K is redshifted compared to the transitions of the other isotopologues and the optically thick transition of CH$_3$OH. This redshifted emission is also seen for the other transitions of CH$_3$OH with very high $E_{\rm up}$ ($>$ 700 K, \href{url}{https://doi.org/10.5281/zenodo.19595973}).  They consequently do not exactly probe the same region as the transitions of the other isotopologues or it could mean that the blueshifted emission of the line is partially absorbed.  

\subsection{Methodology for the radiative transfer modeling}
\label{sect_LTE}

To derive the column densities of the different isotopologues, we assumed LTE. Through non-LTE calculations using the escape probability code RADEX \citep{vanderTak2007}, \citet{Jorgensen2016} demonstrated that LTE was a good approximation for the lines of CH$_3$OH in the PILS spectral range at the densities expected on the scales of the protostars. An analogous study for the COMPASS spectral range yields similar conclusions (see Appendix \ref{app_lte}). Synthetic spectra were produced and compared to the observed line profiles using a Python script, which follows the CASSIS formalism \citep{Bottinelli2024}. In this formalism, the intensity of the line spectrum is calculated as
\begin{equation}
\label{eq:int} 
T_{\rm b} - T_{\rm cont} = \Omega \left( J_{\rm \nu} (T_{\rm ex}) - T_{\rm cont} - J_{\rm \nu} (T_{\rm cmb}) \right) (1 - \exp (-\tau)),
\end{equation}
with 
\begin{equation}
J_{\nu} (T) = \frac{h \nu}{k_{\rm B}} \frac{1}{\exp(h \nu/k_{\rm B} T) - 1} , 
\end{equation}
$\Omega$ as the beam dilution factor, $T_{\rm ex}$ as the excitation temperature of the molecule, $T_{\rm cont}$ as the measured continuum level in K, $T_{\rm cmb}$ as the temperature of the cosmic microwave background (2.73 K), $\nu$ as the frequency, $h$ as the Planck constant, $k_{\rm B}$ as the Boltzmann constant, and 
$\tau$ as the line opacity. We note that $T_{\rm b}$ includes both the line and continuum emission. Through this equation, we assumed that the continuum source fills the beam. This assumption is discussed in Sect. \ref{sect_unc}. 
If the source has a Gaussian emission distribution, the beam dilution factor is equal to 
\begin{equation}
\Omega = \frac{s^2 }{\theta_{\rm maj} \theta_{\rm min}+ s^2},
\end{equation}
with $\theta_{\rm maj}$ and $\theta_{\rm min}$ as the full width at half maximum (FWHM) of the beam major and minor axes, respectively, and $s$ as the source size (i.e., the FWHM of the emission distribution). The opacity of a Gaussian line is described as 
\begin{equation}
\label{eq:tau} 
\tau(\varv) = \tau_0 \exp \left(-\frac{(\varv - \varv_0)^2}{2\sigma_\varv^2} \right),
\end{equation}
with 
\begin{equation}
\sigma_\varv = \frac{\Delta \varv}{2 \sqrt{2 \ln(2)}}.
\end{equation}
The term $\Delta \varv$ represents the FWHM for an optically thin line. $\varv$ is the velocity and $\varv_0$ the velocity at the peak intensity. 
The opacity $\tau_0$ at the peak is equal to
\begin{equation}
\label{eq:tau0} 
\tau_0 = \frac{N~A_{\rm ij}~g_{\rm up}~c^3}{8\pi~\sqrt{2\pi} ~\sigma_\varv~\nu^3~Q(T_{\rm ex})} \left( \exp \left(- \frac{E_{\rm low}}{k T_{\rm ex}} \right)- \exp \left( -\frac{E_{\rm up}}{k T_{\rm ex}} \right) \right),
\end{equation}
with 
$N$ as the column density of the molecule,  
$A_{\rm ij}$ as the Einstein coefficient for spontaneous emission in s$^{-1}$, $g_{\rm up}$ as the degeneracy of the upper state, $c$ as the speed of light, $Q(T_{\rm ex})$ as the partition function for $T$=$T_{\rm ex}$, and $E_{\rm low}$ and $E_{\rm up}$ as the energies of the lower and upper states, respectively. The $E_{\rm up}$, $E_{\rm low}$, $A_{\rm ij}$, and $g_{\rm up}$ values are taken from the spectroscopic databases. The partition function $Q(T)$ is interpolated from the tabulated  values listed in the spectroscopic databases, except for CH$_2$DOH for which we recalculated it (see Appendix \ref{app_spectro}). 
The parameters of column density $N$, excitation temperature $T_{\rm ex}$, FWHM of the line, source velocity $\varv_{\rm LSR}$, and source size $s$ are unknown. The FWHM and $\varv_{\rm LSR}$ can easily be constrained by the line profiles, while the three other parameters require a more careful analysis, as they are interdependent (see Appendix \ref{app_free_parameters}). In our spectral fitting script, the free parameters are constrained using the Levenberg-Marquardt scheme of the Python package lmfit by comparing the synthetic spectra to the observed spectra with a least-squares method. The rms obtained for each spectral window and a 10\% calibration uncertainty were taken into account in the minimization. The velocity interval over which the $\chi^2$ was computed is typically between $-$8 and $-$1 km\,s$^{-1}$, with the exception of lines that are partially blended on the outermost channels, for which we used a smaller range.

The methodology to constrain the free parameters is summarized here. In a preliminary step, we ran scripts with the five parameters free for the detected methanol isotopologues apart from the main one and CH$_3$$^{17}$OH, to get hints on the values of the parameters. The main isotopologue was modeled subsequently, since most of its lines are optically thick and the high excitation transitions may not probe the same region (see Sect. \ref{sect_main_iso}). The lines of CH$_3$$^{17}$OH that would be unblended are too faint and noisy to help constrain the parameters. Its column density was derived using the parameters derived for the other isotopologues.  
The CH$_3$OD $\varv_t$=0 and $\varv_t$=1 transitions, which come from two different CDMS entries, were fitted together. A similar approach was followed for CD$_3$OH. 
For the continuum peak position, we tested with a single velocity component and with two velocity components (corresponding to the blue- and red-shifted emission). A two-component model does not lead to better results due to the high number of unknown parameters. We consequently decide to fit the detected lines with only one component. 
For each isotopologue, we used similar exploration ranges ($s$ between 0.01 and 2$\arcsec$, $N$ between 10$^{13}$ and 10$^{20}$ cm$^{-2}$, $T_{\rm ex}$ between 50 and 300\,K, $\varv_{\rm LSR}$ between -6 and -2 km\,s$^{-1}$, FWHM between 0.5 and 5 km\,s$^{-1}$) and the same initial parameters ($N$ = 10$^{16}$ cm$^{-2}$, $T_{\rm ex}$ = 200\,K, $\varv_{\rm LSR}$ = $-$4.45 km\,s$^{-1}$, FWHM = 2.5 km\,s$^{-1}$). For the source size, we tested different initial values to check that it did not affect the results. 
For $^{13}$CH$_3$OH, CHD$_2$OH, and CH$_3$OD, we found that the derived source sizes (0$\farcs$067--0$\farcs$083) are significantly smaller than the beam size for any initial source size. 
For CD$_3$OH, CH$_2$DOH, and CH$_3$$^{18}$OH, the results depend on the initial source size. The $\chi^2$ is lower for small source sizes (0$\farcs$060 for CD$_3$OH, 0$\farcs$073 for CH$_2$DOH, and 0$\farcs$119 for CH$_3$$^{18}$OH). 
In the case of BHR71IRS1, the beam sizes are relatively similar over the entire frequency range, so it is not the reason why we get small source sizes for some isotopologues. In fact, our sets of lines are usually made of both optically thin and optically thick lines. Given that the source size has a strong impact on the column density, it also has an impact on the line opacities (see Eq. \ref{eq:tau0}) and consequently, on the relative intensities of the lines. The combination of optically thin and somewhat optically thick lines allowed us to obtain an independent constraint on the source size. More detailed explanations are provided in Appendix \ref{app_free_parameters}. The source sizes derived for each isotopologue are discussed in more detail in Sect. \ref{sect_results_isotopo}.
 
Given the respective influences of $N$, $T_{\rm ex}$, and $s$ on each other (see Appendix \ref{app_free_parameters}), it is much better to derive them separately. We therefore proceed as follows: 
\begin{enumerate}
\item We first derived the excitation temperature with both the spectral fitting script and rotational diagrams, only keeping the lines showing the lowest opacities and assuming a beam-filling factor of unity.  
\item We constrained the source size as well as the FWHM and $\varv_{\rm LSR}$ with the spectral fitting script using the previously derived excitation temperature. This time, we used a set of both optically thin and moderately optically thick lines.
\item We ran the spectral fitting script with the previously constrained excitation temperature, source size, velocity, and FWHM to determine the column densities and to calculate the D/H ratios. 
\end{enumerate}

The transitions that are used for the $\chi^2$ minimization are listed on \href{url}{https://doi.org/10.5281/zenodo.19595973}. 
After each run, we checked how the model fits all the transitions. In some rare cases, we included undetected transitions in the $\chi^2$ minimization, as they appeared to be overestimated by some models. This helps to better constrain the parameters and prevents overpredicting these lines. 
We indicated these lines with the symbol $^{**}$ in the tables. In summary, we started with 22, 5, 23, 50, 68, and 44 unblended lines$\footnote{If a line is made up of several transitions blended together, it is only counted once.}$ for $^{13}$CH$_3$OH, CH$_3$$^{18}$OH, CH$_3$OD,  CH$_2$DOH, CHD$_2$OH, and CD$_3$OH, respectively. The small number of lines of CH$_3$$^{18}$OH results from its low column density compared to the other isotopologues. 
The number of lines used in the $\chi^2$ minimization for the final fits, i.e., after removing the highly optically thick lines ($\tau$$\gtrsim$3), is given in Table \ref{tab_fixed_param}. 

\subsection{Less abundant CH$_3$OH isotopologues}
\label{sect_results_isotopo}

\begin{table*}[h!]
\begin{center}
\caption{Column densities (per square centimeter) from the best-fit LTE models for each detected CH$_3$OH isotopologue using a fixed source size, $T_{\rm ex}$, $\varv_{\rm LSR}$, and FWHM.}
\label{tab_fixed_param}
\begin{tabular}{lccccc} 
\hline \hline
 &  Detected lines  & Continuum peak & Redshifted position & Blueshifted position  \\
\hline
Source size (\arcsec) & & 0.07 &  0.07 & 0.07 \\
$T_{\rm ex}$ (K) & &  150  &  140 & 130 \\
$\varv_{\rm LSR}$ (km\,s$^{-1}$) & &    -4.45 &  -3.9 & -4.8 \\
FWHM (km\,s$^{-1}$) & &    2.7  & 2.1 & 2.7 \\
\hline
$N$(CH$_3$OH)$^a$ & &  (5.0$\pm$0.4) $\times$ 10$^{19}$  & (5.2$\pm$0.4) $\times$ 10$^{19}$ & (2.1$\pm$0.2) $\times$ 10$^{19}$ \\
$N$(CH$_3$OH)$^b$ & &  (4.5$\pm$0.8) $\times$ 10$^{19}$ & (4.6$\pm$0.8) $\times$ 10$^{19}$ & (2.2$\pm$0.4) $\times$ 10$^{19}$  \\
$N$(CH$_3$OH)$^c$ & 8 &  (3.5$\pm$0.2) $\times$ 10$^{19}$ & (4.2$\pm$0.3) $\times$ 10$^{19}$  & (3.1$\pm$0.3) $\times$ 10$^{19}$ \\
\hline
$N$($^{13}$CH$_3$OH) & 16 &   (6.6$\pm$0.2) $\times$ 10$^{17}$ & (6.8$\pm$0.3) $\times$ 10$^{17}$ & (3.3$\pm$0.1) $\times$ 10$^{17}$ \\
$N$(CH$_3$$^{18}$OH) & 5 &  (8.9$\pm$0.4) $\times$ 10$^{16}$ & (9.3$\pm$0.5) $\times$ 10$^{16}$ & (3.7$\pm$0.2) $\times$ 10$^{16}$ \\
$N$(CH$_3$$^{17}$OH)$^d$ & 5 &  $\lesssim$(4.1$\pm$0.4) $\times$ 10$^{16}$ & $\lesssim$(3.3$\pm$0.3) $\times$ 10$^{16}$  &  $\lesssim$(2.2$\pm$0.2) $\times$ 10$^{16}$  \\
\hline
$N$(CH$_3$OD) &  23 & (4.2$\pm$0.1) $\times$ 10$^{17}$  & (4.5$\pm$0.2) $\times$ 10$^{17}$ & (1.95$\pm$0.05) $\times$ 10$^{17}$ \\
$N$(CH$_2$DOH) & 46 &  (2.9$\pm$0.6) $\times$ 10$^{18}$   & (3.1$\pm$0.6) $\times$ 10$^{18}$ & (1.6$\pm$0.3) $\times$ 10$^{18}$ \\
$N$(CHD$_2$OH) & 52 &  (1.58$\pm$0.03) $\times$ 10$^{18}$ & (1.78$\pm$0.04) $\times$ 10$^{18}$ & (7.5$\pm$0.2) $\times$ 10$^{17}$ \\
$N$(CD$_3$OH) & 38 & (6.3$\pm$0.2) $\times$ 10$^{17}$ & (7.1$\pm$0.2) $\times$ 10$^{17}$ &  (3.4$\pm$0.1) $\times$ 10$^{17}$ \\
$N$(CD$_3$OD)$^e$ &  &  $<$3 $\times$ 10$^{16}$ & $\lesssim$2 $\times$ 10$^{16}$ &  $<$3 $\times$10$^{16}$ \\ 
\hline
\end{tabular}
\tablefoot{
The reported uncertainties on the fitted isotopologues (1$\sigma$) include the calibration uncertainty and the rms, except for CH$_2$DOH, for which we assumed 20\%. The second column represents the number of detected lines used to determine the column densities in the final fits. 
$^a$ Using the column density of the $^{18}$O isotopologue and a $^{16}$O/$^{18}$O ratio of 557$\pm$30 \citep{Wilson1999}. The uncertainty includes both the uncertainty on the CH$_3^{18}$OH column density and the uncertainty on the $^{16}$O/$^{18}$O ratio. 
$^b$ Using the column density of the $^{13}$C isotopologue and a $^{12}$C/$^{13}$C ratio of 68$\pm$11 \citep{Milam2005}. The uncertainty includes both the uncertainty on the $^{13}$CH$_3$OH column density and the uncertainty on the $^{12}$C/$^{13}$C ratio. 
$^c$ Using the lines of the main isotopologue with the lowest optical depths (<1). The velocity was kept free to better fit these lines that appear redshifted compared to the other isotopologues. For the continuum peak position, the redshifted and the blueshifted positions, we find $\varv_{\rm LSR}$ = -4.0 km\,s$^{-1}$, $\varv_{\rm LSR}$ = -3.6 km\,s$^{-1}$, and $\varv_{\rm LSR}$ = -4.4 km\,s$^{-1}$, respectively. 
$^d$ Given the tentative detection and the low signal-to-noise ratios of the lines, the column densities could be overestimated.
$^e$ 3$\sigma$ upper limits.}
\end{center}
\end{table*}

Models with large source sizes do not reproduce the observations as well as those with smaller source sizes ($<$0$\farcs$1). Therefore, we followed the methodology described in Sect. \ref{sect_LTE} to constrain the excitation temperature and the source size separately. For each isotopologue (except for CH$_3$$^{17}$OH), we selected the lines showing the lowest opacities and we ran the spectral fitting scripts without beam dilution. 
We started with the analysis of the continuum peak position. 
The best-fit excitation temperatures are 134, 79, 136, 144, 144, and 169 K for $^{13}$CH$_3$OH, CH$_3$$^{18}$OH, CH$_3$OD, CH$_2$DOH, CHD$_2$OH, and CD$_3$OH, respectively. A comparison with rotational diagrams is shown in Appendix \ref{sect_RD}. The excitation temperatures of the different isotopologues are in agreement with the ones derived with the spectral fitting script. They range between 141 and 167 K, except for CH$_3$$^{18}$OH, which shows an excitation temperature of 67 K. The excitation temperature of CH$_3$$^{18}$OH relies on only five points and a smaller range of covered $E_{\rm up}$, which makes this estimate less reliable than for the other isotopologues. We consequently adopted the same excitation temperature of 150\,K for all isotopologues in the next steps.

We then ran the models again with the excitation temperature fixed, but this time with all lines. 
We found source sizes of 0$\farcs$081, 0$\farcs$071, 0$\farcs$071, 0$\farcs$075, and 0$\farcs$066 for $^{13}$CH$_3$OH, CH$_3$OD, CH$_2$DOH, CHD$_2$OH, and CD$_3$OH, respectively. 
According to their best-fit models, $^{13}$CH$_3$OH, CH$_2$DOH, CHD$_2$OH, and CD$_3$OH have a significant number of lines with high optical depths ($\tau$$>$3). We reran the code by excluding these lines to not give too much weight to these lines in the $\chi^2$ minimization. The source sizes tend to be slightly smaller, 0$\farcs$067 for $^{13}$CH$_3$OH and CHD$_2$OH, 0$\farcs$061 for CH$_2$DOH and 0$\farcs$057 for CD$_3$OH. The difference in source sizes between the different isotopologues is minor. 
So we used a unique source size of 0$\farcs$07 for all isotopologues for the next runs (see the impact of a slightly different source size in Sect. \ref{sect_unc}). We double-checked by eye that the source size is clearly constrained. For that, we compared the predictions for the source size of 0$\farcs$07 and a source size significantly larger than the beam (no dilution) for each isotopologue.  The models for $^{13}$CH$_3$OH, CH$_2$DOH, CHD$_2$OH, CD$_3$OH, and CH$_3$OD confirm that a small source size is in better agreement with the observations. For CH$_3$$^{18}$OH, the two models are similar except for the faint line at 289.8 GHz ($E_{\rm up}$ = 26\,K and $A_{\rm ij}$ = 4\,$\times$\,10$^{-5}$\,s$^{-1}$), which is better reproduced with a small source size. 

A velocity of $-$4.45 ($\pm$0.1) km\,s$^{-1}$ seems to be appropriate for fitting all the isotopologues. We consequently used this value for all isotopologues. The FWHM ranges from 2.7 to 3\,km\,s$^{-1}$, depending on the isotopologue. We adopted here a FWHM of 2.7\,km\,s$^{-1}$ in agreement with $^{13}$CH$_3$OH, CHD$_2$OH, as well as CH$_3$$^{18}$OH. The uncertainty on this value results from the spectral resolution, which is about 0.5\,km\,s$^{-1}$, the larger line widths of the optically thick lines that depend on the opacity values, and the optically thin lines that are usually faint and consequently noisier than the brighter lines. 

With all these parameters fixed, we derived the column densities using the set of optically thin and moderately optically thick lines ($\tau$$\lesssim$3) defined above. The results obtained for each isotopologue are given in Table \ref{tab_fixed_param}. Comparisons of the observations with the best-fit models obtained with both free and fixed parameters can be found in Fig. \ref{fig_model_13CH3OH} for $^{13}$CH$_3$OH, Fig. \ref{fig_model_CH318OH} for CH$_3$$^{18}$OH, Fig. \ref{fig_model_CH3OD} for CH$_3$OD, Fig. \ref{fig_model_CH2DOH} for CH$_2$DOH, Fig. \ref{fig_model_CHD2OH} for CHD$_2$OH, and Fig. \ref{fig_model_CD3OH} for CD$_3$OH. The similar agreement of the two models with the observations clearly indicates that the chosen values for the model with the fixed parameters are appropriate for each isotopologue. CD$_3$OD is not detected with a 3$\sigma$ upper limit of 3\,$\times$\,10$^{16}$ cm$^{-2}$. The 3$\sigma$ upper limit was determined so that the lines that are predicted to be the brightest ones by the LTE model in the areas of noise (i.e., where there is no blending with any other species) reach 
\begin{equation}
I_{\rm max} = 3~{\rm rms} \sqrt{{d\varv}/{\rm FWHM}},
\end{equation}
with d$\varv$ as the spectral resolution in velocity units. 

To constrain the column density of CH$_3$$^{17}$OH, we assumed the same parameters as for the other isotopologues. We only used the transitions that are found to be unblended. We excluded the lines which are too bright to be reproduced by the model without overpredicting the other lines. Only very faint lines remain (indicated with a blue $*$ in Fig. \ref{fig_model_CH317OH}). Two undetected transitions were included to avoid overestimating the column density. The column density we derived with the spectral fitting code is 4.1\,$\times$\,10$^{16}$ cm$^{-2}$. It leads to a $^{18}$O/$^{17}$O ratio of $\sim$2.1, which is not too far from the isotopic ratio of 3--4 found in the interstellar medium \citep{Zhang2020,Zou2023}. It is possible that the column density is still overestimated. We would need more sensitive observations to derive a precise $^{18}$O/$^{17}$O ratio. 

 A similar analysis with the spectral fitting script was done for the redshifted and blueshifted positions. The excitation temperatures seem to be slightly lower for both positions. We used 140 K for the redshifted position and 130 K for the blueshifted position. A source size of 0$\farcs$07 seems to be in good agreement with the observations in both cases. The FWHM is smaller toward the redshifted position ($\sim$2.1 km\,s$^{-1}$), while it is similar to the continuum peak position for the blueshifted position ($\sim$2.7 km\,s$^{-1}$). The $\varv_{\rm LSR}$ is about $-$3.9 and $-$4.8 km\,s$^{-1}$ for the redshifted and blueshifted positions, respectively. The column densities obtained with these parameters fixed are included in Table \ref{tab_fixed_param}. In the end, with the same source size for the three positions, we find that the column densities are similar at the continuum peak position and the redshifted one, while they are a factor of two lower at the blueshifted position. We note that the $^{18}$O/$^{17}$O ratio is in better agreement with the interstellar value for the redshifted position (2.8$\pm$0.3), where the lines are slightly clearer than at the other two positions. This confirms the high uncertainty on the derived column densities of CH$_3$$^{17}$OH.

\subsection{Main isotopologue}
\label{sect_main_iso}

The lines of the main isotopologue are usually optically thick, which makes it difficult to derive precise column densities.
When possible, its column density is derived using the less abundant isotopologues $^{13}$CH$_3$OH and CH$_3$$^{18}$OH and assuming a $^{12}$C/$^{13}$C ratio of 68$\pm$11 \citep{Milam2005}  and a $^{16}$O/$^{18}$O ratio of 557$\pm$30 \citep{Wilson1999}, respectively. For the continuum peak position, this leads to a column density of methanol of (5.0$\pm$0.4)\,$\times$\,10$^{19}$ cm$^{-2}$ using the $^{18}$O isotopologue and (4.5$\pm$0.8)\,$\times$\,10$^{19}$ cm$^{-2}$ using the $^{13}$C isotopologue. The two values are in agreement with each other, if we consider the uncertainties from the spectral fitting and from the $^{12}$C/$^{13}$C and $^{16}$O/$^{18}$O isotopic ratios. The column densities of methanol obtained with the two methods at the other positions are also in agreement.

Based on the results obtained with the low abundant isotopologues, there are eight CH$_3$OH lines in the COMPASS spectral range with opacities lower than 1. They are relatively noisy, but we can use them to constrain the column density directly from the main isotopologue (see Table \ref{tab_fixed_param}). It is clear, especially from the lines with higher optical depths (1--3), that there is a systematic velocity shift between the predictions and the observations, if we use the same velocities as for the other isotopologues. So we kept the $\varv_{LSR}$ free to constrain the column densities. This shift in velocity by about 0.4 km\,s$^{-1}$ is consistent with what we have seen in the moment 1 maps (Fig. \ref{fig_maps_moment1}). The column densities are listed in Table \ref{tab_fixed_param}. If we include the seven lines with opacities between 1 and 3, the column density is a factor of two lower. The optically thin lines count less in the $\chi^2$ minimization, as they are noisier. For this reason, we report the value obtained with only the eight lines with $\tau$$<$1. The results for the blueshifted position are more uncertain than for the other positions due to noisier lines, which could explain the 40$-$48\% higher column density found with the main isotopologue compared to the estimates from $^{13}$C- and $^{18}$O-bearing isotopologues. For the other two positions, the column density of the main isotopologue is 12$-$30\% lower than the ones obtained with the less abundant isotopologues. But we should keep in mind that the different velocity of the main isotopologue could suggest a partial absorption on the blue side of the lines or that a different part of the envelope is probed. So this does not necessarily point to different isotopic ratios than the standard values. 
For this reason, we use the value obtained from CH$_3$$^{18}$OH for the column density of methanol for the rest of the paper. 

\subsection{D/H ratios}

\begin{table*}[ht!]
\begin{center}
\caption{Statistically corrected D/H ratios (as percentages) of CH$_3$OH and column density ratios of CH$_2$DOH/CH$_3$OD in BHR71IRS1.}
\label{tab_comparison_positions}
\begin{tabular}{lccc} 
\hline\hline
  & Continuum peak & Redshifted position & Blueshifted position \\
\hline 
 & \multicolumn{3}{c}{Statistically corrected D/H ratios of the hydroxyl group (in \%)} \\ 
\hline
CH$_3$OD/CH$_3$OH & 0.8$\pm$0.2 & 0.9$\pm$0.2 & 0.9$\pm$0.2 \\
CD$_3$OD/CD$_3$OH & $<$5 & $\lesssim$3 & $<$9 \\
\hline 
 & \multicolumn{3}{c}{Statistically corrected D/H ratios of the methyl group (in \%)} \\ 
\hline
CH$_2$DOH/CH$_3$OH & 1.9$\pm$0.6 & 2.0$\pm$0.6 & 2.5$\pm$0.7 \\
CHD$_2$OH/CH$_3$OH & 10$\pm$2 & 11$\pm$2 & 11$\pm$2 \\
CD$_3$OH/CH$_3$OH  & 23$\pm$5 & 24$\pm$5 & 25$\pm$5 \\
CD$_3$OD/CH$_3$OD  &  $<$42 & $\lesssim$35 & $<$53 \\
CHD$_2$OH/CH$_2$DOH & 55$\pm$17 & 58$\pm$17 & 48$\pm$15 \\
CD$_3$OH/CHD$_2$OH  & 119$\pm$24 & 121$\pm$24 & 135$\pm$27 \\
\hline
 & \multicolumn{3}{c}{Column density ratios of CH$_2$DOH/CH$_3$OD} \\ 
\hline
CH$_2$DOH/CH$_3$OD & 7$\pm$2 & 7$\pm$2 & 8$\pm$3 \\
\hline
\end{tabular}
\tablefoot{We indicate conservative uncertainties of 20\% for all ratios, except those involving CH$_2$DOH, for which we use 30\% (see Sect. \ref{sect_unc}).} 
\end{center}
\end{table*}

To derive the D/H ratios, we needed to statistically correct the column density ratios based on the following equations (for more details, see Appendix B in \citealt{Manigand2019}):
\begin{equation}
\rm {\it N}(CH_3OD)/{\it N}(CH_3OH) = (D/H),
\end{equation}
\begin{equation}
\rm {\it N}(CD_3OD)/{\it N}(CD_3OH) = (D/H), 
\end{equation}
\begin{equation}
\rm {\it N}(CH_2DOH)/{\it N}(CH_3OH) = 3 \times (D/H),
\end{equation}
\begin{equation}
\rm {\it N}(CHD_2OH)/{\it N}(CH_3OH) = 3 \times (D/H)^2,
\end{equation}
\begin{equation}
\rm {\it N}(CD_3OH)/N(CH_3OH) = (D/H)^3,
\end{equation}
\begin{equation}
\rm {\it N}(CD_3OD)/{\it N}(CH_3OD) = (D/H)^3,
\end{equation}
\begin{equation}
\rm {\it N}(CHD_2OH)/{\it N}(CH_2DOH) = (D/H),
\end{equation}
and
\begin{equation}
\rm {\it N}(CD_3OH)/{\it N}(CHD_2OH) = (D/H) / 3.
\end{equation}
The derived D/H ratios at the three positions are presented in Table \ref{tab_comparison_positions}. For any given isotopologue pair, the measured ratios are in good agreement across the three positions. 
From these calculations, we can see that the D/H ratio is a factor of two higher on the methyl group ($\sim$1.9\%) than on the hydroxyl group ($\sim$0.8\%). This corresponds to a column density ratio of CH$_2$DOH/CH$_3$OD of 7$\pm$2. The degree of deuteration increases with the number of D atoms on the methyl group (from 2\% for CH$_2$DOH to 23\% for CD$_3$OH).  The upper limit obtained for the deuteration of the hydroxyl group from CD$_3$OD/CD$_3$OH ($\lesssim$3\%) is consistent with the D/H ratio obtained from CH$_3$OD/CH$_3$OH ($\sim$0.8\%). Similarly, the upper limit obtained for the deuteration of the methyl group from CD$_3$OD/CH$_3$OD ($\lesssim$35\%)
is in agreement with the D/H ratio derived from CD$_3$OH/CH$_3$OH ($\sim$23\%).

\subsection{Uncertainties}
\label{sect_unc}

The uncertainties on the column densities obtained with the $\chi^2$ minimization contain the rms and the calibration uncertainty (10\%) of the observations. They are about $\sim$ 2--4\% (1$\sigma$) for all isotopologues except for CH$_3$$^{17}$OH ($\sim$10\%), as indicated in Table \ref{tab_fixed_param}. However, these uncertainties assume that the values of the fixed parameters ($s$, $T_{\rm ex}$, FWHM) are necessarily exact. We discuss the impact that a change in source size, excitation temperature, and FWHM would have on the column densities and isotopic ratios. We also consider the impact that an error on the continuum subtraction would have on the results. The uncertainty on the continuum subtraction is typically about 1 rms \citep{Plunkett2026}. 
Given the uncertainty in the intrinsic line intensities of CH$_2$DOH mentioned in the spectroscopic studies (\citealt{Pearson2012,Oyama2023}, see Appendix \ref{app_spectro}), we used an uncertainty of 20\% for the column density of this isotopologue rather than the smaller value derived with the spectral fitting procedure. 

For the source size, $T_{\rm ex}$, and FWHM, we used the range of values found for the different isotopologues to investigate how it affects the column densities. 
The FWHM is uncertain by 0.3 km\,s$^{-1}$, which results in a difference of 0.3--7\% on the column densities depending on the isotopologue. 
For the excitation temperature, a change by $\pm$20 K leads to column densities that differ by $\sim$10--20\%, except for CH$_3$$^{18}$OH, for which it is smaller (7\%) and CH$_3$OH, for which it is higher (50\%). For the source size, using 0$\farcs$06 or 0$\farcs$08 leads to variations of $\sim$30--40\%.
 If we consider that the continuum is 1 rms above or below the value we estimated, the line intensity changes and it results in a difference of 20--30\% for the column densities of the different isotopologues. 

Regarding the statistically corrected D/H ratios, we can calculate their uncertainties by propagating those of the column densities presented in Table \ref{tab_fixed_param}. This leads to uncertainties of 4--8\% for all ratios except those involving CH$_2$DOH that are much higher (26\% for CH$_2$DOH/CH$_3$OH and 22\% for CHD$_2$OH/CH$_2$DOH) due to the higher uncertainty of 20\% we used for the CH$_2$DOH column density. 
We can explore how the choices of the parameters and the continuum level impact the D/H ratios, similarly to what we have done for the column densities. 
The same assumptions for the fixed parameters and the same continuum level were considered for the two species when deriving the ratios. Then, we compared their variations with the uncertainties calculated by propagating the uncertainties on the column densities presented in Table \ref{tab_fixed_param}.
We first note that using data with a continuum level different by $\pm$1 rms has a very minor effect on the D/H ratios ($\lesssim$2\%). The changes are below the 1$\sigma$ uncertainties estimated with the propagation of uncertainties for each D/H ratio. A similar conclusion is obtained for the FWHM ($\sim$0.2--7\%) and the source size ($\sim$2--8\%). They are all within the derived 1$\sigma$ uncertainties, except the one obtained from CD$_3$OH/CH$_2$DOH, which is within 2$\sigma$. 
The main changes are seen for an excitation temperature that varies by $\pm$20 K. The variation of the D/H ratios derived from CH$_2$DOH/CH$_3$OH (17\%), CHD$_2$OH/CH$_3$OH (5\%), and CHD$_2$OH/CH$_2$DOH (9\%) correspond to less than 1$\sigma$, while the other D/H ratios from CH$_3$OD/CH$_3$OH, CD$_3$OH/CH$_3$OH, and CD$_3$OH/CHD$_2$OH change within 2$\sigma$ ($\sim$10\%,  6\%, and 9\%, respectively). 
A possible variation in all fixed parameters would lead to changes in the D/H ratios of 6--18\% for all ratios. 
Consequently, if we consider uncertainties of 20\% on all ratios except those involving CH$_2$DOH, they should be conservative. For the ratios with CH$_2$DOH, an uncertainty of 30\% should be more appropriate considering the higher values of their propagated uncertainties. This also applies to the CH$_2$DOH/CH$_3$OD column density ratio. These values are adopted for the uncertainties listed in Table \ref{tab_comparison_positions}.

Our analysis assumes that the continuum emission fully fills the beam (Equation \ref{eq:int}). If the continuum emission were more compact ($\Omega$ $\leq$ $\Omega_d$ $<$ 1 with $\Omega_d$ the beam dilution factor for the continuum), the continuum seen by the gas would be higher due to its dilution in the beam. In other words, $T_{\rm cont}$ would become $T_{\rm cont}/\Omega_d$ on the right-hand side of Equation \ref{eq:int}. 
However, the continuum cannot be as compact as the emitting region of the CH$_3$OH isotopologues, as this would cause their spectral lines to appear in absorption rather than emission (i.e., $T_{\rm cont}/\Omega_d$ would be significantly higher than $J_{\rm \nu} (T_{\rm ex})$ for the excitation temperatures found in BHR71IRS1, $\sim$130--150 K).  We infer that the continuum size should be $\gtrsim$0.3$\arcsec$ for the lines to remain in emission, which is only slightly smaller than the angular resolution of our observations. With a continuum size similar to the beam size (i.e., a dilution factor of 0.5), the column densities would be underestimated by a factor of $\sim$1.5--2.4. The ratios should not be significantly affected.

\section{Discussion}
\label{sect_discussion}

\subsection{Deuterium fractionation ratios of methanol in protostars}

\begin{table*}[h!]
\begin{center}
\caption{Comparison of the statistically corrected D/H ratios of CH$_3$OH (in \%) and the column density ratio of CH$_2$DOH/CH$_3$OD in BHR71IRS1 (continuum peak) with other protostars.}
\label{tab_comparison_protostars}
\begin{tabular}{lcccccc} 
 \hline \hline
Source  & BHR71IRS1 & IRAS16293 A & IRAS16293 B & HOPS 373SW & B335$^a$ & V883-Ori \\ 
\hline
& \multicolumn{5}{c}{Statistically corrected D/H ratios of the hydroxyl group (in \%) } \\
\hline
CH$_3$OD/CH$_3$OH &  0.8$\pm$0.2      	&	& 0.32$\pm$0.09   & 			& 			& 0.10$\pm$0.02\\
CD$_3$OD/CD$_3$OH &   $<$5    		 	&	& $\lesssim$6        & 			& 			&                \\
\hline
& \multicolumn{5}{c}{Statistically corrected D/H ratios of the methyl group (in \%) } \\
\hline
CH$_2$DOH/CH$_3$OH  &  1.9$\pm$0.6     	& 2.8$\pm$1.2       	& 1.8$\pm$0.5 	& 3.9$\pm$1.4   	& 			&  0.042$\pm$0.009 \\
CHD$_2$OH/CH$_3$OH  & 10$\pm$2      & 7.5$\pm$1.1        & 7.7$\pm$1.2 	& 17$\pm$4	& 			& $<$0.6 \\
CD$_3$OH/CH$_3$OH     & 23$\pm$5     &                            	& 15$\pm$2 		& 26$\pm$4  \\
CD$_3$OD/CH$_3$OD     & $ <$42		        &                            	&  $\lesssim$39 	\\

CHD$_2$OH/CH$_2$DOH  & 55$\pm$17       &  20$\pm$7         	& 33$\pm$9 		& 75$\pm$28 & 14--29 		& $<$10 \\
CD$_3$OH/CHD$_2$OH   & 119$\pm$24       	&                        	& 52$\pm$15 	& 61$\pm$27    \\
\hline
 & \multicolumn{5}{c}{Column density ratios of CH$_2$DOH/CH$_3$OD} \\ 
\hline
CH$_2$DOH/CH$_3$OD & 7$\pm$2 &  & 16$\pm$5 & & & 1.22$\pm$0.02 \\
\hline
\end{tabular}
\tablefoot{$^a$ We did not include the ratios with respect to CH$_3$OH due to optical depth issues \citep{Okoda2024}. References: IRAS16293~A and B: \citet{Jorgensen2018}, \citet{Manigand2020}, \citet{Drozdovskaya2022}, \citet{Ilyushin2022,Ilyushin2023,Ilyushin2024}. We recalculated the D/H ratios of IRAS16293~B following the change of column density of CH$_2$DOH (see Appendix \ref{app_spectro}). We assumed an uncertainty of 20\% for the column density of CD$_3$OH. HOPS 373SW: \citet{Lee2023}. For the CH$_3$OH abundance, we used here $^{13}$CH$_3$OH and a $^{12}$C/$^{13}$C ratio of 68.  B335: \citet{Okoda2024}. V883-Ori: \citet{Zeng2025}. We calculated the D/H ratios using the CH$_3^{18}$OH column density from \citet{Jeong2024}.}
\end{center}
\end{table*}

\citet{Yang2020} previously studied the chemical composition of BHR71IRS1 with a resolution of $\sim$0$\farcs$32. That paper derived column densities for CH$_3$OH, $^{13}$CH$_3$OH, and CH$_2$DOH based on a small number of lines (1, 2, and 3, respectively) and assuming a source size of 0$\farcs$5. The excitation temperature was estimated to be $\sim$100 K based on the CH$_2$DOH lines. Using $^{13}$CH$_3$OH and a $^{12}$C/$^{13}$C ratio of 68, it leads to a statistically corrected D/H ratio of 2.6\%, which is relatively close to the value we derived in this study (1.9$\pm$0.6\% at the continuum peak).

\begin{figure}[h!]
\begin{center}
\includegraphics[width=9cm]{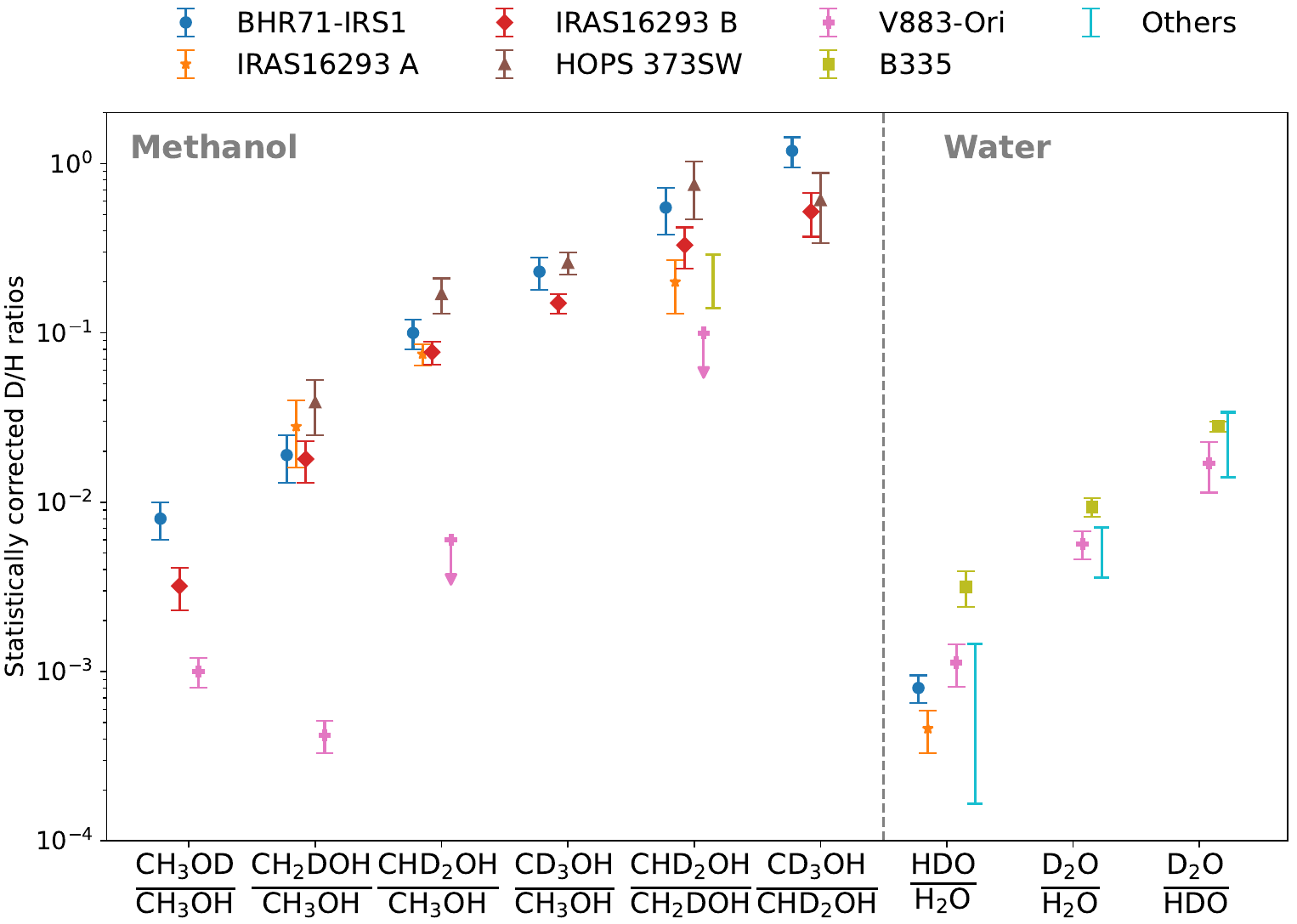}
\caption{Comparison of the statistically corrected D/H ratios of methanol measured in several low-mass protostars. For BHR71IRS1, we only indicate the D/H ratios at the continuum peak position since they are all consistent for the three positions. A range was used for the CHD$_2$OH/CH$_2$DOH value in B335, as the value varies depending on the radius. The right part of the plot shows the statistically corrected D/H ratios of water obtained with interferometers from the same set of sources along with the range of values reported in the literature for additional low-mass protostars \citep{Persson2013,Persson2014,Coutens2014,Jensen2019,Jensen2021b,Andreu2023,Tobin2023,Leemker2025}.}
\label{fig_compar_DH} 
\end{center}
\end{figure}

The methanol D/H ratios of BHR71IRS1 are compared with other protostars (see Table \ref{tab_comparison_protostars} and Fig. \ref{fig_compar_DH}). For a relevant comparison between the sources, we only included results obtained with interferometers. As the lines of the main isotopologue are usually optically thick \citep{Bianchi2020,Cruz2025}, we also only used the column densities of methanol obtained from studies of the $^{18}$O or $^{13}$C isotopologues, as we did for BHR71IRS1. 
To avoid any confusion, we also systematically used statistically corrected D/H ratios in this section.

Numerous estimates of the CH$_2$DOH/CH$_3$OH ratio are available in the literature. However, these studies often rely on a limited number of spectral lines, which can significantly affect the derived column densities due to uncertainties in the CH$_2$DOH line strengths (S$\mu^2$, see Appendix \ref{app_spectro}). As a result, comparisons between different studies should be approached with caution, particularly since they are not based on the same sets of transitions. 
In general, the D/H ratio in BHR71IRS1 (1.9$\pm$0.6\%) is similar to most protostars. For example, there are estimates of $\sim$2--4\% for HH212 \citep{Bianchi2017,Leec2019}, $\sim$1.7\% for L1551 IRS5 \citep{Cruz2025}, $\sim$2.8\% for B1-c, and $\sim$1.4\% for Serpens S68N \citep{vanGelder2022}.  In the B component of IRAS16293, \citet{Jorgensen2018} found a D/H ratio of 2.4\%, which would give 1.8\% with the recalculation of the partition function of CH$_2$DOH presented in Appendix \ref{app_spectro}. \citet{Manigand2020} find 2.8\% for IRAS16293~A. Slightly lower values are found in the isolated Class 0 protostar L483 ($\sim$0.8\%, \citealt{Jacobsen2019}) and the Class I protostar SVS13-A ($\sim$0.7\%, \citealt{Bianchi2017}), but this difference may not be significant. In five hot corinos of Orion studied by \citet{Hsu2022}, the D/H ratios are about 0.7--1.7\%. 
The only source that seems to be less deuterated is the protoplanetary disk V883-Ori, for which \citet{Yamato2024} and \citet{Jeong2024} derived a D/H ratio of 0.2\%, i.e., an order of magnitude lower than the BHR71IRS1 value. Using the recent laboratory spectroscopic data for CH$_2$DOH from \citet{Oyama2023}, \citet{Zeng2025} found a D/H ratio of 0.7\%, but it is based on the column density of the main isotopologue. If we use the column density of CH$_3$$^{18}$OH \citep{Jeong2024} and a $^{16}$O/$^{18}$O ratio of 557, the D/H ratio of V883-Ori would be about 0.04\%, which would confirm the lower deuteration level of this source. 

The D/H ratio from CH$_3$OD has been estimated in a few sources \citep[e.g.,][]{Taquet2019}. However, these studies relied on spectroscopic data that was not publicly available and the partition functions presented large variations between studies. The recent spectroscopic work by \citet{Ilyushin2024} is now included in the CDMS database. Their reanalysis of IRAS16293 B reveals that the column density previously reported by \citet{Jorgensen2018} differs by a factor of about five. As a result, and to ensure consistency in the comparisons, we included only the results obtained using the same spectroscopic dataset, i.e., for IRAS16293 B \citep{Ilyushin2024} and for V883-Ori \citep{Zeng2025}. 
Similarly, for CHD$_2$OH and CD$_3$OH, we only focused on the works using the most recent spectroscopic data available on CDMS. Doubly deuterated methanol was studied toward the two components of IRAS16293 by \citet{Drozdovskaya2022}, HOPS 373SW by \citet{Lee2023}, B335 by \citet{Okoda2024}, and V883-Ori by \citet{Zeng2025}, while triply deuterated methanol was investigated toward IRAS16293 B \citep{Ilyushin2022}. An upper limit for CD$_3$OD is also available in \citet{Ilyushin2023} for the same source. 

Table \ref{tab_comparison_protostars} summarizes the measured D/H ratios of BHR71IRS1, IRAS16293 A and B, HOPS 373W, B335, as well as V883-Ori. For B335, we excluded the ratios with respect to CH$_3$OH as its column density relies on the main isotopologue \citep{Okoda2024}. So we only considered its CHD$_2$OH/CH$_2$DOH ratio. For the CH$_3$OH column density of V883-Ori, we used the value of CH$_3$$^{18}$OH derived by \citet{Jeong2024} multiplied by 557. A comparison of these D/H ratios can be found in Fig. \ref{fig_compar_DH}. The D/H ratios of BHR71IRS1 measured on the methyl group are similar to the other sources, except for V883-Ori, which would be less deuterated. For CH$_3$OD/CH$_3$OH, the comparison is more difficult due to a lack of measurements. BHR71IRS1 seems to be more deuterated on the hydroxyl group than IRAS16293 B and V883-Ori. 

The origin of the variations in the deuteration between the OH and CH$_3$ groups is unclear (see the discussion in \citealt{Zeng2025}).
The D/H ratio is higher on the CH$_3$ group than on the OH group for both BHR71IRS1 and IRAS16293 B. Interestingly, the trend is opposite for V883-Ori. While methanol is expected to form by hydrogenation of CO ices, several experiments indicate that H--D substitution reactions (abstraction of a H atom followed by the addition of a D atom) 
lead to an efficient formation of CH$_2$DOH, CHD$_2$OH, and CD$_3$OH on grain surfaces \citep{Nagaoka2005,Nagaoka2007,Hidaka2009}. However, such substitutions do not occur on the hydroxyl group, as \citet{Nagaoka2005} did not find any CH$_3$OD after exposure of CH$_3$OH to D atoms. As hypothesized in \citet{Kulterer2022}, the CH$_3$OD abundance would be low in the prestellar stage and its formation would mainly take place in the protostellar stage, but this seems to be contradicted by a recent detection of CH$_3$OD in a prestellar core \citep{Kulterer2026}. Experiments have also shown that rapid H--D exchanges between water and methanol occur on the hydroxyl group at 120 K and above \citep{Ratajczak2009}, which seems to be in agreement with observations of Orion KL \citep{Wilkins2022}. After implementing these exchanges in a chemical model, \citet{Faure2015} found that the CH$_2$DOH/CH$_3$OD ratio would scale inversely with the D/H ratio of water and that it would reproduce relatively well the observations. This means that the CH$_3$OD/CH$_3$OH ratio would tend to equilibrate with the water deuteration, which is usually known to be lower than in COMs. A more quantitative comparison is presented in Sect. \ref{sect_water}. 
The efficiency of H--D substitution reactions in the methyl group of methanol at low temperatures associated with the rapid H--D exchange reactions between the hydroxyl group of methanol and water at higher temperatures can explain the lower D/H ratio on the OH group than on the CH$_3$ group in BHR71IRS1 and IRAS16293 B. 

A clear trend of increasing deuteration with the number of D atoms in the methyl group is observed for BHR71IRS1, as well as for the other sources. A similar pattern is also seen in other molecules, where multiple deuterated forms have been detected (H$_2$O: \citealt{Coutens2014}, CH$_3$CN: \citealt{Calcutt2018}, HCOOCH$_3$: \citealt{Manigand2019}, CH$_3$CHO: \citealt{Ferrer2023}, CH$_3$OCH$_3$: \citealt{Richard2021}, c-C$_2$H$_4$O: \citealt{Muller2023}). This suggests that once a molecule becomes deuterated, it is more likely to undergo further deuteration, possibly due to favorable reaction pathways, increased stability of partially deuterated species, or different formation timescales. In the case of methanol, the H--D substitution reactions on the methyl group could possibly play a key role.

While the CH$_3$OD/CH$_3$OH, CD$_3$OH/CH$_3$OH, and CD$_3$OH/CHD$_2$OH ratios of methanol in BHR71IRS1 are higher compared to IRAS16293 B, the other D/H ratios are similar to components A and B of IRAS16293 within the uncertainties. 
The D/H ratios of methanol measured for HOPS~373SW \citep{Lee2023}, which is located in Orion in a more clustered region than BHR71IRS1, are also relatively similar. It is consequently not possible to conclude on an overall higher deuteration of methanol for the isolated source BHR71IRS1.  This could mean that, contrary to water, the environment of low-mass protostars does not have an impact on the methanol deuteration. Measurements in a larger sample of sources with similar datasets are needed to confirm the absence of trend between the methanol deuteration and the source environment. Future comparisons with the other sources of the COMPASS Large Program will be important to answer this question.

\subsection{Comparison with the deuterium fractionation ratio of water}
\label{sect_water}

Water deuteration was studied in the protostar BHR71IRS1 by \citet{Jensen2019}. 
They found an HDO/H$_2$O ratio of (1.8$\pm$0.4)\,$\times$\,10$^{-3}$ in its warm inner region using ALMA observations. This means that the water D/H ratio (0.9\,$\times$\,10$^{-3}$) is a factor of 10 and 20 lower than the statistically corrected CH$_3$OD/CH$_3$OH and CH$_2$DOH/CH$_3$OH ratios, respectively. Finding a higher D/H ratio in COMs than in water is very common \citep[e.g.,][] {Parise2003,Coutens2012,Persson2013,Jorgensen2018}. This would be explained by an earlier formation of water compared to CH$_3$OH. Indeed, water can efficiently form during the diffuse or translucent phases, whereas methanol requires colder temperatures and higher densities, conditions under which CO can freeze out onto dust grains. As explained by \citet{Furuya2016} and \citet{Jensen2021a}, water and its deuterated counterparts would form in two distinct stages. The first stage occurs in the molecular cloud phase. Deuteration is inefficient due to the moderately high temperatures and low densities, which mainly promotes the formation of nondeuterated water. The second stage takes place in the denser and colder phase. Under these conditions, water can form, but not as efficiently as previously. However, deuteration is significantly more efficient. Consequently, most of the deuterated water forms in this second stage. The substantial amount of nondeuterated water produced during the first stage ultimately results in a lower overall HDO/H$_2$O ratio.
 
Doubly deuterated water has not been studied in this source, but it was found in two other isolated protostars (L483 and B335), as well as in the clustered source NGC1333~IRAS2A that the D$_2$O/HDO column density ratio is higher than the HDO/H$_2$O column density ratio by a factor of two to seven  \citep{Coutens2014,Jensen2021b}. The statistically corrected D/H ratio from D$_2$O/H$_2$O is consequently three to five times higher than the one from HDO/H$_2$O.
From CH$_2$DOH/CH$_3$OH to CHD$_2$OH/CH$_3$OH, the D/H ratio increases by a factor of about five, which is relatively similar to water. This could indicate that the deuteration of water and methanol occur simultaneously. Nevertheless, D$_2$O observations toward BHR71IRS1 would be necessary to conclude. 

\citet{Taquet2014} predicted D/H ratios for both methanol and water in protostellar envelopes using a grain multi-layer approach and a 1D dynamical model. The addition, substitution, and abstraction reactions studied by \citet{Hidaka2009} are included in their chemical network. The predicted D/H ratios for HDO and D$_2$O\footnote{We statistically corrected all the ratios presented in their Table 7.} are (0.32--1.2)\,$\times$\,10$^{-2}$ and (0.6--2.5)\,$\times$\,10$^{-2}$, respectively, toward the inner protostellar region. Even though the predicted values for HDO are slightly higher than the observations, the values for D$_2$O are in agreement (see Fig. \ref{fig_compar_DH}). However, for methanol, the trend is different. The predicted D/H ratios of CH$_2$DOH ((0.2--1.8)\,$\times$\,10$^{-2}$) and CH$_3$OD ((0.3--3.5)\,$\times$\,10$^{-2}$) reproduce the observations relatively well, but the predictions for the CH$_2$DOH/CH$_3$OD ratio ($\sim$1.5--1.7) fail to reproduce the high abundance ratios measured toward BHR71IRS1 and IRAS16293B. The D/H ratios for CHD$_2$OH and CD$_3$OH are also underestimated by typically an order of magnitude. The disagreement for CH$_2$DOH/CH$_3$OD could be explained by the fact that the rapid H--D exchange reactions between the hydroxyl group of CH$_3$OH and H$_2$O \citep{Faure2015} are not taken into account. However, the reason for the low CHD$_2$OH/CH$_3$OH and CD$_3$OH/CH$_3$OH ratios of the chemical model is not clear.
 
\citet{Faure2015} showed that the range of CH$_2$DOH/CH$_3$OD ratios observed in protostars can be explained by rapid H--D exchange reactions between water and the hydroxyl group of CH$_3$OH at elevated temperatures. Using the constraints derived for both methanol and water in BHR71IRS1, we can test whether this mechanism accounts for the observations. As described in \citet{Faure2015}, the CH$_2$DOH/CH$_3$OD abundance ratio should follow
\begin{equation}
\frac{\rm CH_2DOH}{\rm CH_3OD} = 3 \alpha_m \times 2 \exp(-50/T_d) \times \frac{\rm H_2O}{\rm HDO}
\label{eq_faure},
\end{equation}
with $T_d$ as the dust temperature ($\sim$100 K) and $\alpha_m$ as the accreting D/H ratio during methanol formation.
The statistically corrected D/H ratio for CH$_2$DOH/CH$_3$OH is used as a proxy for $\alpha_m$. 
With an HDO/H$_2$O ratio of 1.8\,$\times$\,10$^{-3}$ and $\alpha_m$ = 1.9\%, the CH$_2$DOH/CH$_3$OD ratio should be around 38, which is roughly five times higher than the CH$_2$DOH/CH$_3$OD ratio measured in BHR71IRS1 (7$\pm$2). However, as discussed above, the water D/H in the grain mantles is not homogeneous. The innermost layers, formed early during the star formation process, are expected to be water-rich but weakly deuterated, whereas the outer layers should be more strongly enriched in deuterium. Since methanol forms only after CO depletion onto grains, it is likely under-abundant in the innermost layers. 
Consequently, the regions where methanol is present (and where H--D exchanges with water can occur) should correspond to a higher level of water deuteration. 
If we instead assume an HDO/H$_2$O ratio of about 1\%, the predicted CH$_2$DOH/CH$_3$OD ratio decreases to $\sim$7, in agreement with the observations. Such a level of water deuteration is consistent with the high D$_2$O/HDO ratios measured in protostars \citep{Coutens2014,Jensen2021b}. If a protostar in a clustered environment and an isolated one exhibit similar CH$_2$DOH/CH$_3$OH ratios, the one in the clustered region is expected to have a CH$_2$DOH/CH$_3$OD ratio higher by a factor of $\gtrsim$2 compared to the isolated source, as its water deuteration should be lower by a similar factor $\gtrsim$2 \citep{Jensen2019}. This expectation appears to be consistent with the value measured toward IRAS16293B (16$\pm$5). 
In V883-Ori, the CH$_2$DOH/CH$_3$OD ratio is significantly lower ($\sim$1.2). 
Using the above expression and $\alpha_m$ = 0.042\%, the expected ratio is 0.76 for the measured HDO/H$_2$O ratio of 2\,$\times$\,10$^{-3}$ and 0.15 considering an HDO/H$_2$O ratio of 1\%. Although these values are lower than the observed ratio of 1.2, the overall trend is reproduced. More advanced gas-grain chemical models of protostellar objects, incorporating the processes described in \citet{Faure2015}, would help refine this interpretation.

\subsection{Comparison of the gas and ice CH$_3$OH/H$_2$O abundance ratios}

The CH$_3$OH/H$_2$O abundance ratios in ices have been estimated toward several low-mass star-forming regions thanks to infrared telescopes such as the Very Large Telescope (VLT), the Spitzer Space Telescope, and more recently the James Webb Space Telescope (JWST). These values range from 2 to $\sim$32\% \citep[e.g.,][]{Pontoppidan2004,Boogert2015,Perotti2020,Chen2024,Rocha2025}. A higher value of $\sim$66\% was found for SVS 4--9, but the determination of the local continuum is uncertain for this source \citep{Perotti2020}. 

Using the water results from \citet{Jensen2019}, we can estimate the CH$_3$OH/H$_2$O abundance ratio in the gas phase. 
\citet{Jensen2019} estimated a H$_2$$^{18}$O column density of (2.0$\pm$0.4)\,$\times$\,10$^{15}$ cm$^{-2}$ assuming no beam dilution. If we correct for the beam dilution assuming the same emission size as methanol (0$\farcs$07) and assuming a $^{16}$O/$^{18}$O ratio of 557, the column density of water becomes (1.4$\pm$0.3)\,$\times$\,10$^{20}$ cm$^{-2}$. This leads to a gas-phase CH$_3$OH/H$_2$O ratio of 34$\pm$8\%, which is consistent with the upper range of the ice values. A similar ratio is expected if CH$_3$OH forms on the grains and then sublimates in the hot corino (without destruction of methanol or water in the gas phase). JWST observations of ices toward BHR71IRS1 and its surrounding environment will place the gas-phase CH$_3$OH/H$_2$O ratio in the context of its icy origin in a future work.

\section{Conclusions}
\label{sect_conclusion}

In this paper, we have studied the isotopic fractionation of methanol toward the low-mass protostar BHR71IRS1 using a large spectral survey between 279 and 312 GHz obtained within the ALMA Large Program COMPASS. Using LTE models, we determined the column densities of all the isotopologues available in the public spectroscopic databases. The source size was found to be small ($\sim$0$\farcs$07), and the excitation temperature is about 150~K at the continuum peak position. Using the column density of CH$_3$$^{18}$OH and applying a standard isotopic $^{16}$O/$^{18}$O ratio of 557$\pm$30, we found a CH$_3$OH column density of (5.0$\pm$0.4)\,$\times$\,10$^{19}$\,cm$^{-2}$, which is in agreement (within the uncertainties) with the value obtained from $^{13}$CH$_3$OH, and a $^{12}$C/$^{13}$C ratio of 68$\pm$11. A tentative detection has been obtained for CH$_3$$^{17}$OH. 

The detections of CH$_3$OD, CH$_2$DOH, CHD$_2$OH, and CD$_3$OH allowed us to constrain the deuteration of methanol in this source. We find that the deuteration for the monodeuterated isotopologues is higher on the methyl group (D/H = 1.9$\pm$0.6\%) than on the hydroxyl group (D/H = 0.8$\pm$0.2\%),  corresponding to a CH$_2$DOH/CH$_3$OD column density ratio of 7$\pm$2. The deuteration increases with the number of D atoms in the methyl group and reaches 23$\pm$5\% for the statistically corrected D/H ratio of CD$_3$OH/CH$_3$OH.
These ratios and trends are relatively similar to previous results for other protostars, such as IRAS16293 and HOPS 373SW. However, they seem to differ from the eruptive young star V883-Ori, where methanol would be less deuterated. Further studies on methanol deuteration will be carried out for the protostars observed as part of COMPASS. The results obtained from datasets covering the same large spectral range should allow us to better understand the drivers of methanol deuteration. 

\section*{Data availability}
The lists of the selected unblended transitions for each isotopologue are available at \href{url}{https://doi.org/10.5281/zenodo.19595973}.

\begin{acknowledgements}
We thank the anonymous referee for helpful comments that improved this manuscript. 
This paper makes use of the following ALMA data: ADS/JAO.ALMA\#2022.1.00316.L. ALMA is a partnership of ESO (representing its member states), NSF (USA) and NINS (Japan), together with NRC (Canada), NSTC and ASIAA (Taiwan), and KASI (Republic of Korea), in cooperation with the Republic of Chile. The Joint ALMA Observatory is operated by ESO, AUI/NRAO and NAOJ. The project leading to this publication has received support from ORP, that is funded by the European Union's Horizon 2020 research and innovation programme under grant agreement No 101004719 [ORP].
We acknowledge financial support from the following: 
the European Research Council (ERC) under the European Union's Horizon 2020 research and innovation programme (Grant agreement No. 949278, Chemtrip; A.C., F.C.-S.M., A.A., P.M.), 
the National Research Foundation of Korea (NRF), funded by the Korea government (MSIT; grant numbers RS-2024-00416859 and RS-2026-25490557; J.-E.L., J.-H.J., H.-S.Y., C.-H.K.), 
the Basic Science Research Program through the NRF of Korea, funded by the Ministry of Education (grant number RS-2025-25402622; J.-H.J.), 
the Ministry of Education of Taiwan (a grant from Center for Informatics and Computation in Astronomy, CICA, and grant number 110J0353I9) and the National Science and Technology Council, Taiwan (grant NSTC111-2112-M-007-014-MY3, NSTC113-2639-M-A49-002-ASP, and NSTC113-2112-M-007-027; D.H.), 
the Thematic Action ``Physique et Chimie du Milieu Interstellaire'' (PCMI) of INSU Programme National ``Astro'', with contributions from CNRS Physique \& CNRS Chimie, CEA, and CNES  (S.M.), 
the RIKEN Special Postdoctoral Researcher Program (Fellowships; J.F.A., S.Z.), 
the NKFIH NKKP grant ADVANCED 149943 and the NKFIH excellence grant TKP2021-NKTA-64, the Ministry of Culture and Innovation of Hungary from the National Research, Development and Innovation Fund, financed under the NKKP ADVANCED funding scheme (Project no.149943) as well as COST (European Cooperation in Science and Technology) through COST Action PLANETS CA22133 (A.K), 
the ESO and IAU Gruber Foundation Fellowship programs (P.N.), 
the Max Planck Society (S.S.), 
a Grant-in-Aid from the Ministry of Education, Culture, Sports, Science, and Technology of Japan (20H05845, 20H05844), and a pioneering project in RIKEN (Evolution of Matter in the Universe) (Y.-L.Y.), 
the National Radio Astronomy Observatory and the North American ALMA Science Center (M.R, Z.T.).  
\end{acknowledgements}

\bibliographystyle{aa} 
\bibliography{Biblio} 

\begin{appendix}

\section{Constraining parameters with LTE models}
\label{app_free_parameters}

As explained in Sect. \ref{sect_LTE}, by comparing observed and synthetic spectra produced with LTE models, we can constrain the following parameters: $N$, $T_{\rm ex}$,  FWHM, $\varv_{\rm LSR}$, and $s$. In this section, we discuss from a theoretical point of view how these parameters are interrelated and how they can be constrained. 
The $\varv_{\rm LSR}$ is the velocity at the peak of intensity of a Gaussian line. It is the easiest parameter that can be constrained, as it is not impacted by the other parameters. It can easily be derived by eye, but a code based on Gaussian fits should give a similar result. 
The FWHM is also relatively easy to constrain as long as the lines are optically thin. When a line is optically thick, it loses its Gaussian profile and its width 
increases due to the saturation of its peak. In case of moderately optically thick lines, the LTE spectral fitting code should take into account the broadening of the lines due to the line opacities to derive the FWHM. Since the opacities depend on the emitting source size and excitation temperature, the determination of the FWHM should be done when these two parameters are well constrained. 
The interplay between the three other parameters is more complex. If the source size is known, constraining the column density and the excitation temperature is not a problem, as long as there is a decent number of lines covering a large range of $E_{\rm up}$. If the source size is unknown, we can consider three different cases: i) we only have optically thin lines, ii) we only have optically thick lines, and iii) we have a mix of optically thin and optically thick lines.

\subsection{Only optically thin lines}
\label{sect_opt_thin}

In case of optically thin lines ($\tau << 1$), Eq. \ref{eq:int} becomes
\begin{equation}
T_b - T_{\rm cont} =   
~\Omega~(J_\nu (T_{\rm ex}) - T_{\rm cont} - J_\nu(T_{\rm cmb}))~\tau.
\end{equation}
For a given $T_{\rm ex}$ and a given source size, the line intensity at the peak (i.e., $T_b - T_{\rm cont}$  for $\varv$ = $\varv_0$) is proportional to $\tau_0$, and consequently to $N$, based on Eq. \ref{eq:tau0}.
The higher the column density, the higher the line intensity. 

When several optically thin transitions are observed with beam sizes that differ across transitions, it is easy to constrain the source size if the emission size is smaller than the beam sizes. Indeed, in this case, the beam dilution factor $\Omega$ should vary depending on the transitions and, consequently, affect the line intensities differently. 

However, if the beam sizes are similar, there is a degeneracy between the source size and the column density ($T_b - T_{\rm cont} \propto \Omega N_{\rm tot}$). The column density increases when the source size decreases. In this case, it is not possible to constrain the source size, and a size needs to be assumed to derive a column density.

\subsection{Only optically thick lines}

In case of optically thick lines ($\tau >> 1$), at the velocity peak, Eq. \ref{eq:int} becomes
\begin{equation}
T_b - T_{\rm cont} =   
~\Omega~(J_\nu (T_{\rm ex}) - T_{\rm cont} - J_\nu(T_{\rm cmb})).
\label{eq_opt_thick_lines}
\end{equation}
This means that, once the column density is sufficiently high, the line intensity saturates. Consequently, if all the lines we observe are optically thick, it is only possible to derive a lower limit on the column density. Obviously, this value depends on the assumed source size and excitation temperature. If there exists no constraint on the source size (e.g., no existing high resolution maps, no constraint based on less abundant isotopologues or related species in the same source), the lower limit on the column density should be derived assuming no beam dilution. 

\subsection{Both optically thin and thick lines}

If the molecule has a significant number of both optically thin and moderately optically thick lines, we can constrain the source size. 
If the beam sizes vary, this is the same as the case with only optically thin lines (see Sect. \ref{sect_opt_thin}). If the beam sizes are similar for each transition, we can still derive a source size although this method is probably not as precise as the case with different beam sizes. 
As explained above, for optically thin lines, $T_b - T_{\rm cont}$ increases with $N$ while, for optically thick lines, it saturates at a certain level.
 This means that if we assume a source size that is much larger than its actual value, the opacities are underestimated, and the model predictions tend to overpredict the optically thick lines or underpredict the optically thin ones. On the contrary, if we assume a source size that is much smaller, the opacities are overestimated and some of the normally optically thin lines may not be reproduced due to the saturation of their line intensities.  
 Therefore, it should be possible to constrain the source size. 
 
When considering both the excitation temperature and the source size as free parameters, we usually observe a degeneracy between these two parameters that can lead to an uncertain source size. Ideally, to avoid any bias, it is preferable to first constrain the excitation temperature based on the optically thin lines only. Once $T_{\rm ex}$ is known, we can derive the source size and the column density by using both the moderately optically thick and optically thin lines. The difficulty here lies in the selection of the optically thin lines. Indeed, the opacities depend on the chosen parameters for $N$, $T_{\rm ex}$, and $s$. As these parameters are initially not constrained, it is important to check how the opacities change with the parameters and only select the transitions with the lower opacities in a relative way. For example, if no beam dilution is considered while the source is smaller than the beam size, the opacities are underestimated. So a much lower threshold than one is required. Several iterations with a critical approach may be needed for a proper selection of the optically thin lines. 

It is important to note that the limitation of this approach is the use of a 0D model with a single component of a certain size and temperature, which is a  simplified representation of the inner region of a protostar (e.g., temperature and density gradient, disk structure). Consequently, while such models can give hints on the source size, we should still be careful not to overinterpret the results given by automated line fits, especially in the case of a very low number of lines and/or very small uncertainties given by the code. Comparing the results for both a small source size and a large source size by eye is recommended to see how many lines are used for the constraint. If only one or two lines are better fitted with a small source size, a likely alternative is that these lines are blended with unidentified species, for example. 
 It should also be noted that the reduced $\chi^2$ method assumes that the free parameters are independent, which is not the case here. 
 This is the reason why we carried out a separate study of the sensitivity to estimate the parameter uncertainties (see Sect. \ref{sect_unc}).

\section{Spectroscopic data}
\label{app_spectro}

The spectroscopic data used in this study all come from the CDMS database, except for CH$_2$DOH. 
The entry of CH$_3$OH is based on \citet{Xu2008}. The method and program used to reduce the data is described in \citet{Xu1995a} and \citet{Xu1995b}. 
The entry of $^{13}$CH$_3$OH is based on a list of lines published in \citet{Xu1997} with the methodology described in \citet{Xu1996}. 
CH$_3$$^{18}$OH is investigated in \citet{Fisher2007} with data coming from \citet{Hughes1951,Gerry1976,Hoshino1996,Predoi-Cross1997} and \citet{Ikeda1998}. The entry of CH$_3$OD is based on \citet{Ilyushin2024}. 
CHD$_2$OH is studied in \citet{Drozdovskaya2022}. The calculations are based on \citet{Coudert2021} and most of the data come from this study.  
Most of the data of CD$_3$OH and CD$_3$OD are from \citet{Ilyushin2022} and \citet{Ilyushin2023}, respectively. The entry of CH$_3$$^{17}$OH is based on the work by \citet{Muller2024}. 
Given the excitation temperatures found here, no correction for additional contributions in the partition function should be needed for any of these isotopologues. 

The spectroscopy of CH$_2$DOH available in the JPL database comes from \citet{Pearson2012}. It is mentioned that the a-type transitions ($\Delta$K$_a$ = 0) are more reliable than the b-type and c-type transitions ($\Delta$K$_c$ = $\pm$1) for the determination of the column densities. Unfortunately, we have only one a-type transition among the 50 transitions used in the $\chi^2$ minimization and it has a high optical depth. \citet{Oyama2023} recently measured the CH$_2$DOH transitions in the spectral range 216--264 GHz in the laboratory. They found that, in their study, the S$\mu^2$ is systematically higher for the a-type J=5$-$4 transitions than in JPL ($\sim$13$-$27\%), while it varies for the b-type and c-type transitions: some transitions show higher S$\mu^2$, while others show lower S$\mu^2$ within a factor of two. Since we have a large number of transitions in our spectral survey, we expect that this mitigates the possible errors in the determination of the column density. It is clear that some lines are overestimated, while others are underestimated in Fig. \ref{fig_model_CH2DOH}. But the column densities do not seem affected significantly when a large number of lines are used, as shown in \citet{Zeng2025}. 
For the partition function, we use a list of energy levels that contains excited torsional states, and rotational energies for high $\varv_t$ values are included up to 1500 cm$^{-1}$ \citep{Spezzano2026}. This corresponds to a small correction factor in comparison to the partition function published on JPL (1.006 at 150\,K and 1.1 at 300\,K). In the PILS study of IRAS16293~B by \citet{Jorgensen2018}, a correction factor to the partition function of 1.457 was applied for an excitation temperature of 300\,K. For a better comparison between the sources, the column density of CH$_2$DOH published for IRAS16293~B is here divided by 1.32 (i.e., 1.457/1.1). For IRAS16293~A, no factor was applied given the lower excitation temperature of CH$_2$DOH (130\,K, \citealt{Manigand2019}). 

\section{LTE assumption}
\label{app_lte}

\begin{figure}[t!]
\begin{center}
\includegraphics[width=8.5cm]{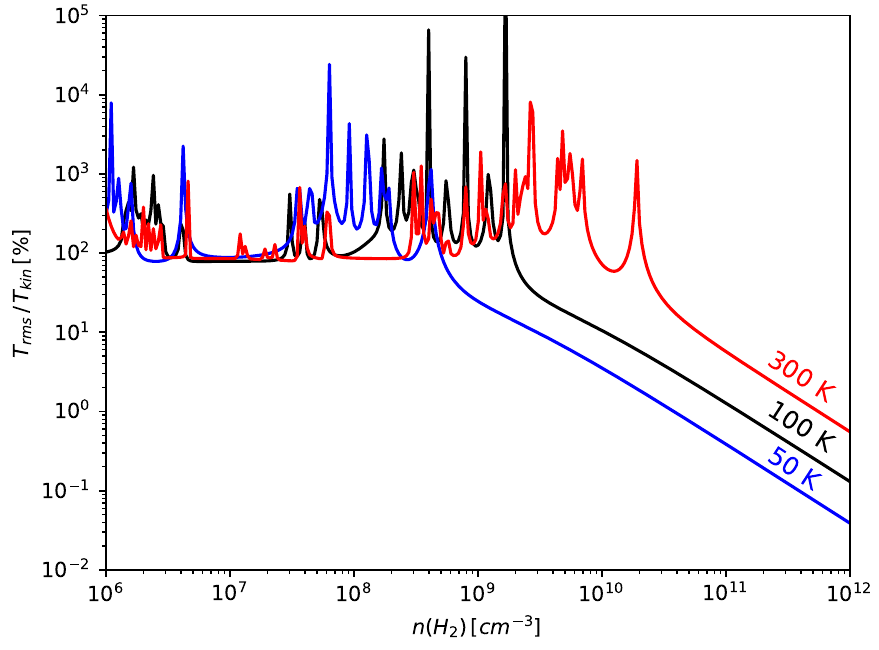}
\caption{Root mean square deviation ($T_{\rm rms}$) relative to the kinetic temperature for all the CH$_3$OH transitions in the 279-312 GHz range as a function of density for a sufficiently low column density ensuring that all transitions are optically thin. The lines in blue, black, and red correspond to kinetic temperatures of 50, 100, and 300 K, respectively.}
\label{fig_ch3oh_lte}
\end{center}
\end{figure}

Similarly to \citet{Jorgensen2016}, we used the RADEX code \citep{vanderTak2007} to check the validity of the LTE approximation for CH$_3$OH, using the  collisional rates determined by \citet{Rabli2010}. The excitation temperatures of all the CH$_3$OH lines in the COMPASS spectral range (279--312 GHz) were calculated for a range of densities between 10$^6$ and 10$^{12}$ cm$^{-3}$ and for three different kinetic temperatures ($T_{\rm kin}$ = 50, 100, and 300 K). The root mean square deviation between the excitation temperatures and the kinetic temperature is defined as 
\begin{equation}
T_{\rm rms} = \sqrt{\frac{1}{N} \sum_i (T_{\rm ex,i} - T_{\rm kin})^2},
\end{equation}
with $i$ referring to each transition and $N$ the total number of CH$_3$OH transitions in the spectral range. 
By plotting $T_{\rm rms}$/$T_{\rm kin}$ as function of density (see Fig. \ref{fig_ch3oh_lte}), we can determine if the LTE assumption holds. In LTE, the excitation temperature should be equal to the kinetic temperature and, as a consequence, a low value is expected for $T_{\rm rms}$/$T_{\rm kin}$. 
For low densities, $T_{\rm rms}$/$T_{\rm kin}$ is relatively high (100\%), and peaks produced by transitions masing appear at specific densities (e.g., \citealt{Jeong2026}). However, from $\gtrsim$10$^9$ cm$^{-3}$, $T_{\rm rms}$/$T_{\rm kin}$ decreases significantly with density. For the high densities found in the inner regions of protostars ($\gtrsim$3\,$\times$\,10$^{10}$ cm$^{-3}$), LTE seems to be a reasonable assumption. 

\section{Rotational diagrams} 
\label{sect_RD}

\begin{figure*}[h!]
\begin{center}
\begin{subfigure}{.33\textwidth}
  \centering
  \includegraphics[width=1.0\linewidth]{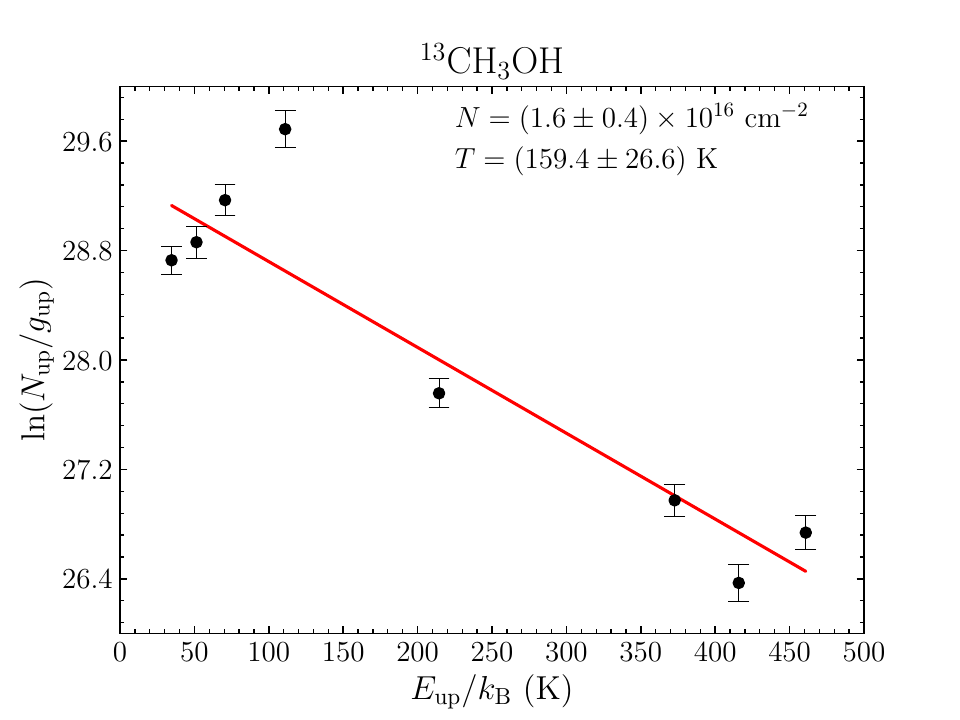}
\end{subfigure}
\begin{subfigure}{.33\textwidth}
  \centering
  \includegraphics[width=1.0\linewidth]{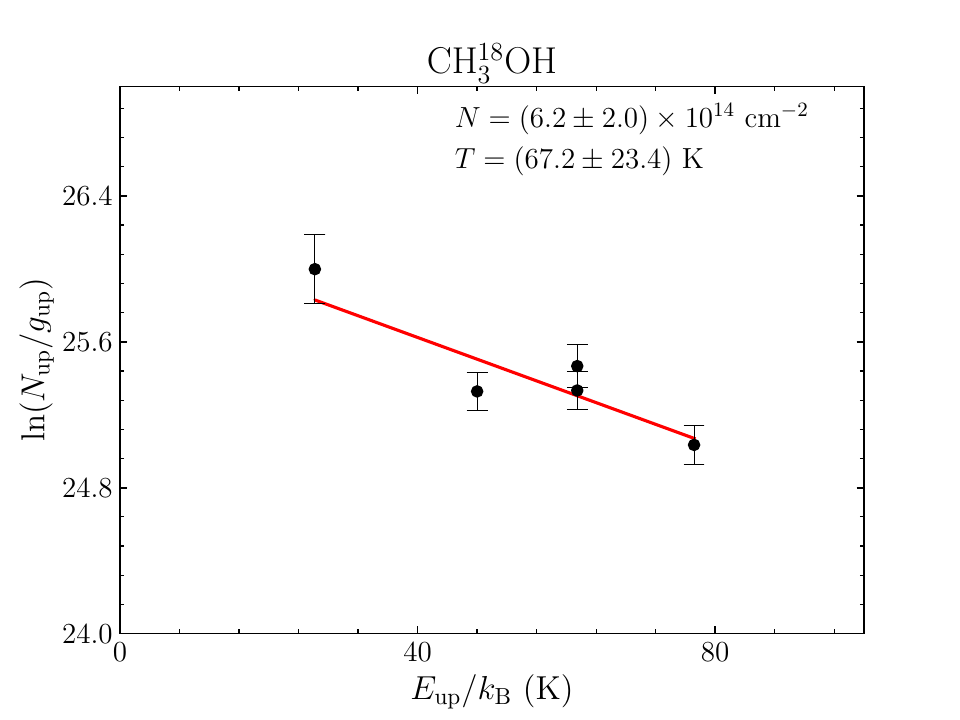}
\end{subfigure}
\begin{subfigure}{.33\textwidth}
  \centering
  \includegraphics[width=1.0\linewidth]{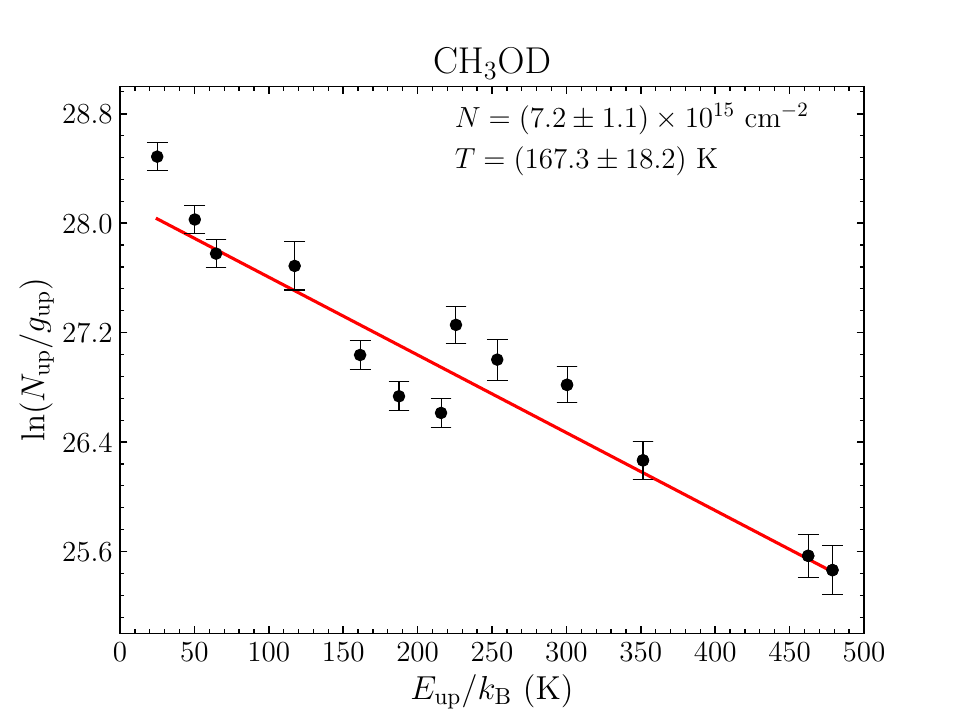}
\end{subfigure}
\vspace{-0.3cm}
\begin{subfigure}{.33\textwidth}
  \centering
 \includegraphics[width=1.0\linewidth]{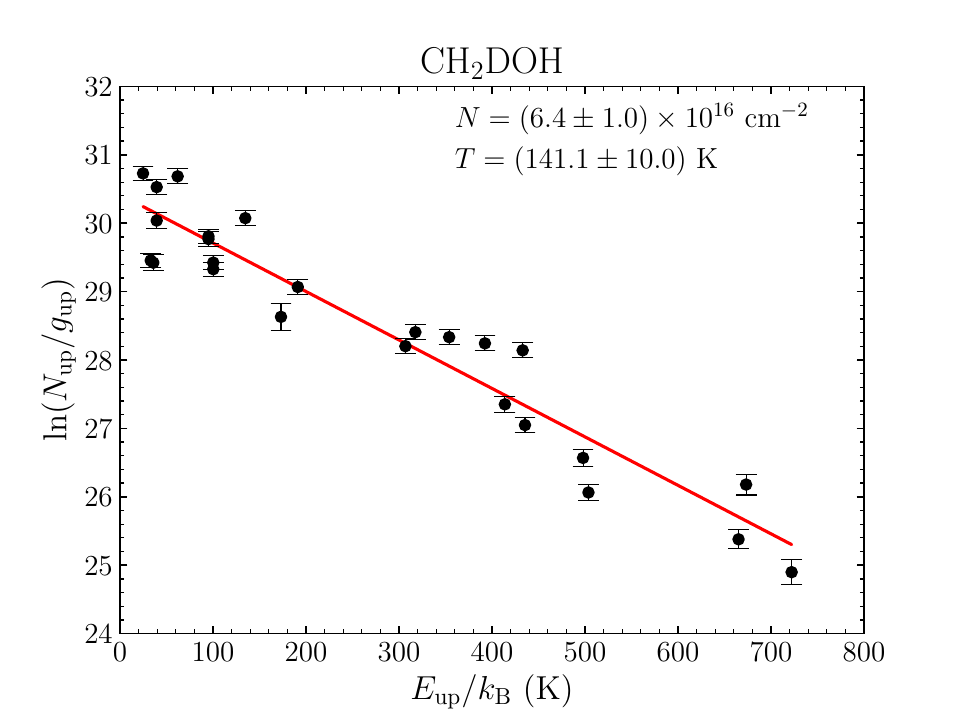}
\end{subfigure}
\begin{subfigure}{.33\textwidth}
  \centering
  \includegraphics[width=1.0\linewidth]{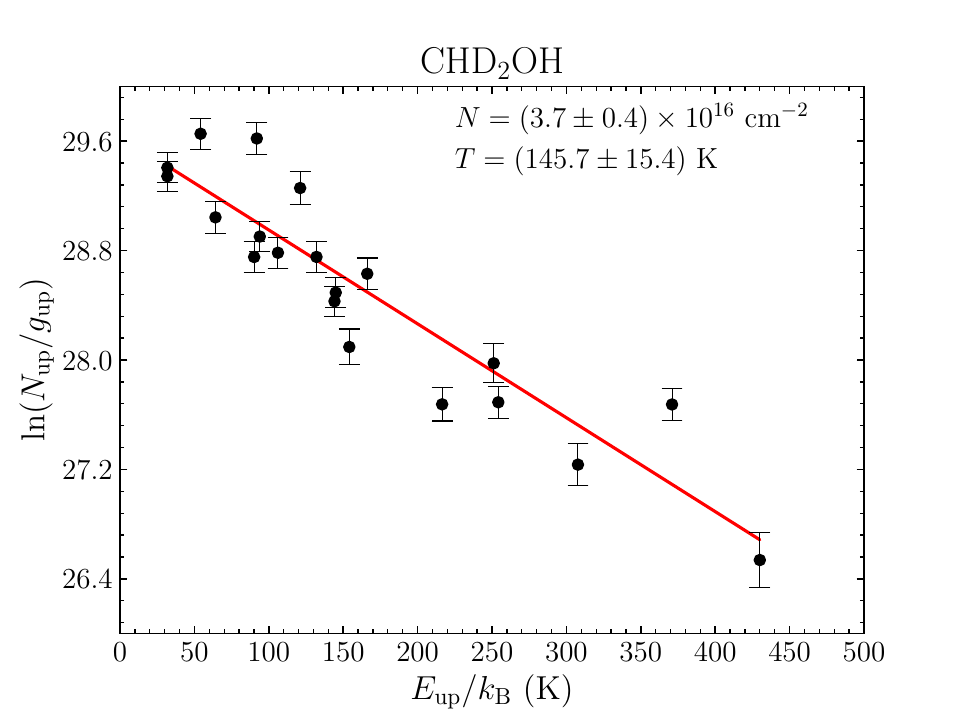}
\end{subfigure}
\begin{subfigure}{.33\textwidth}
  \centering
  \includegraphics[width=1.0\linewidth]{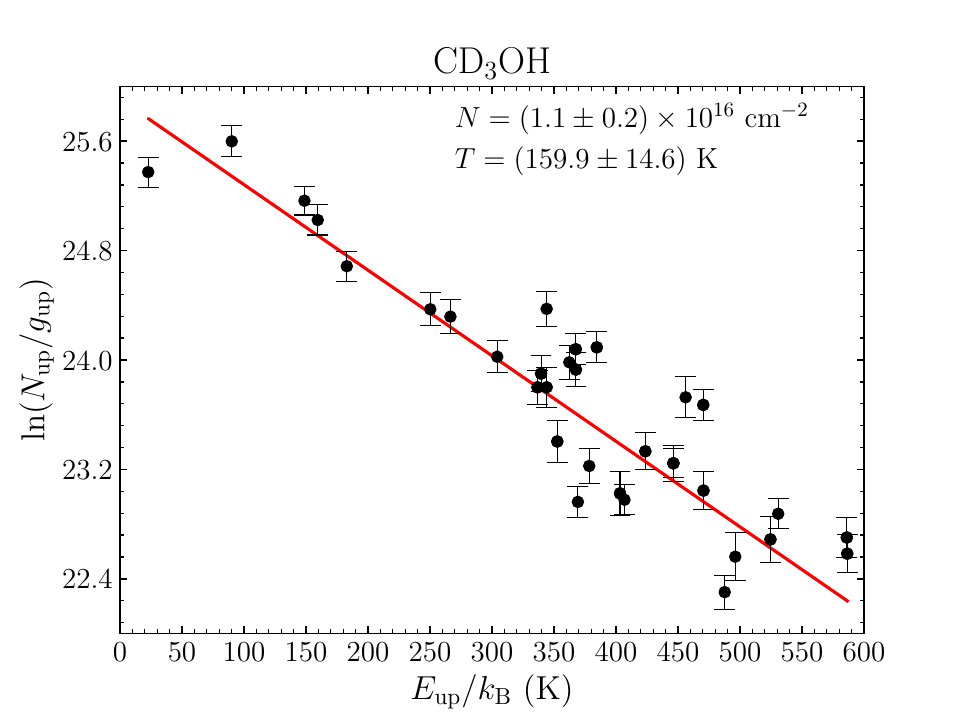}
\end{subfigure}
\caption{Rotational diagrams of the isotopologues of CH$_3$OH after selecting the transitions showing the lowest opacities and without beam dilution. The linear fits (in red) to the data points lead to the listed column densities and excitation temperatures. The column densities are underestimated given that we do not take into account any beam dilution and that the continuum is not negligible in our case. For CH$_3$OD and CD$_3$OH, we included both the $\varv_t$=0 and $\varv_t$=1 transitions (separate CDMS entries). The uncertainties come from the flux uncertainties, which include the calibration uncertainty (10\%) and the noise. }
\label{fig_RD}
\end{center}
\end{figure*}

In Fig. \ref{fig_RD}, we plotted rotational diagrams \citep{Goldsmith1999b} in order to constrain the excitation temperatures and compare them with the values derived with the LTE spectral fitting script in the case of the continuum peak position (see Sect. \ref{sect_results_isotopo}). As the source size is not known at this step, it is not possible to correct for the opacities. For this reason, we only include the lines presenting the lowest opacities for each isotopologue and assume no beam dilution. 
A disadvantage of using rotational diagrams is that we cannot include the lines that are made of several transitions with different $E_{\rm up}$ values. These lines are consequently excluded, while they are kept in the $\chi^2$ minimization with the spectral fitting script. It is important to note that the contribution of the dust continuum is not taken into account in the rotational diagram method. For the inner regions of sources such as BHR71IRS1, the dust continuum is clearly not negligible compared to the excitation temperature. If the continuum is constant, it should have no impact on the slope and consequently on the estimates of the excitation temperatures. However, if the beam sizes vary significantly across the spectral range of the observations, it can lead to large uncertainties on the excitation temperatures, as discussed in \citet{Belloche2019}. In our case, the continuum varies between 36 and 52\,K. The fact that the continuum is not negligible also leads to an underestimation of the column densities listed in Fig. \ref{fig_RD}. 

\FloatBarrier

\section{Spectra of the methanol isotopologues}
\label{sect_spectra}

\begin{figure*}[h!]
\begin{center}
\includegraphics[width=16cm]{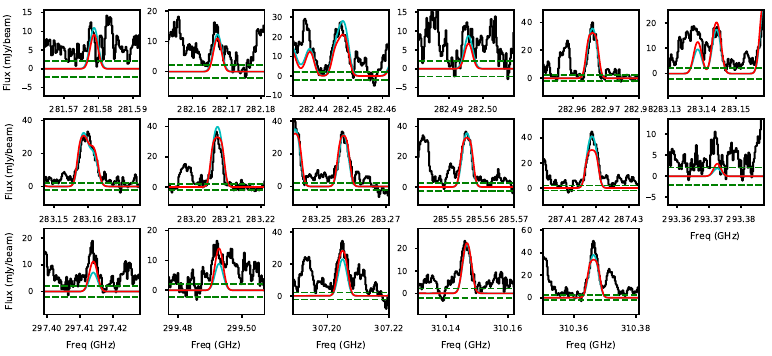}
\caption{Observations of the $^{13}$CH$_3$OH transitions at the continuum peak position (in black) compared to the best-fit models when fixing the parameters $s$, $T_{\rm ex}$, FWHM, and $\varv_{\rm LSR}$ in red and with free parameters in cyan. The dashed green lines correspond to $\pm$1 rms (per channel).}
\label{fig_model_13CH3OH}
\end{center}
\end{figure*}

\FloatBarrier

\begin{figure*}[h!]
\begin{center}
\includegraphics[width=16cm]{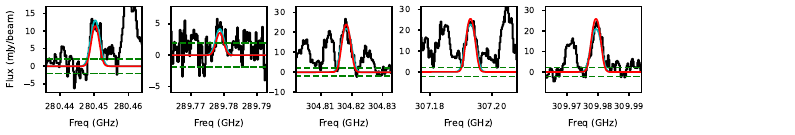}
\caption{Same as Fig. \ref{fig_model_13CH3OH}, but for CH$_3$$^{18}$OH.}
\label{fig_model_CH318OH}
\end{center}
\end{figure*}

\begin{figure*}[h!]
\begin{center}
\includegraphics[width=16cm]{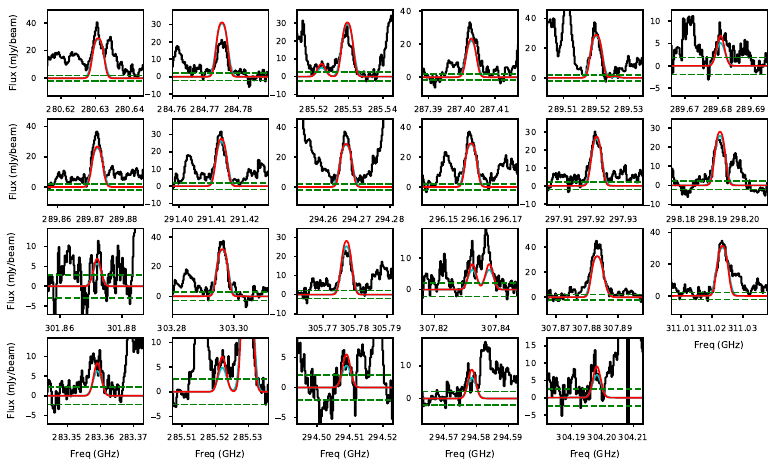}
\caption{Same as Fig. \ref{fig_model_13CH3OH}, but for CH$_3$OD.}
\label{fig_model_CH3OD}
\end{center}
\end{figure*}

\begin{figure*}[h!]
\begin{center}
\includegraphics[width=16cm]{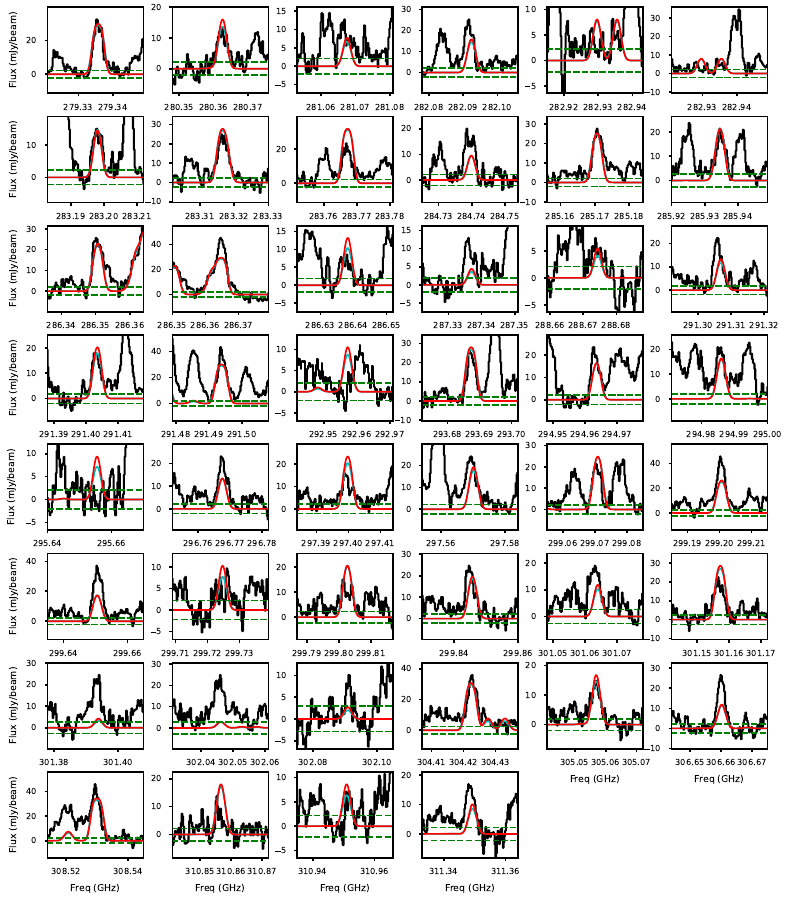}
\caption{Same as Fig. \ref{fig_model_13CH3OH}, but for CH$_2$DOH.}
\label{fig_model_CH2DOH}
\end{center}
\end{figure*}

\begin{figure*}[h!]
\begin{center}
\includegraphics[width=16cm]{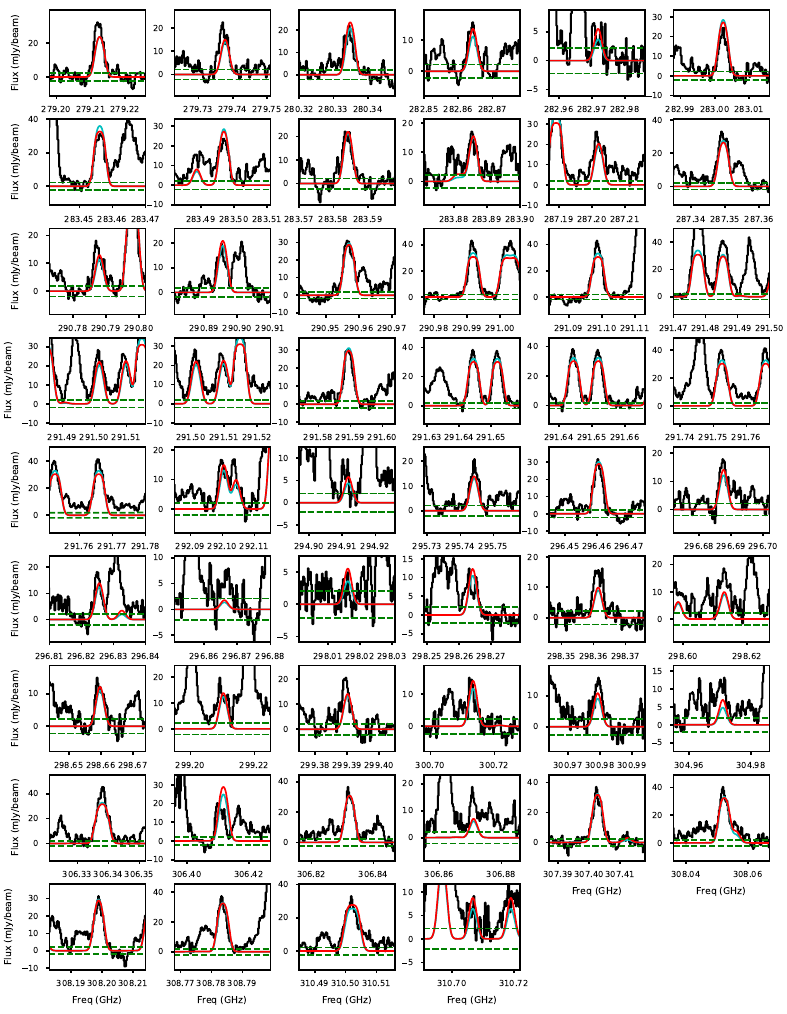}
\caption{Same as Fig. \ref{fig_model_13CH3OH}, but for CHD$_2$OH.}
\label{fig_model_CHD2OH}
\end{center}
\end{figure*}

\begin{figure*}[h!]
\begin{center}
\includegraphics[width=16cm]{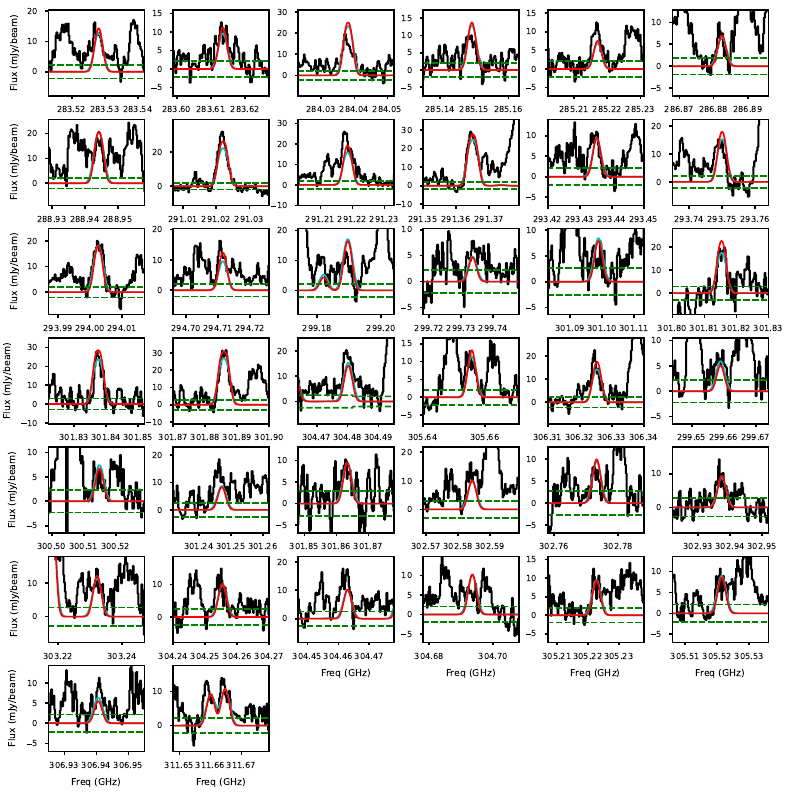}
\caption{Same as Fig. \ref{fig_model_13CH3OH}, but for CD$_3$OH.}
\label{fig_model_CD3OH}
\end{center}
\end{figure*}

\begin{figure*}[h!]
\begin{center}
\includegraphics[width=16cm]{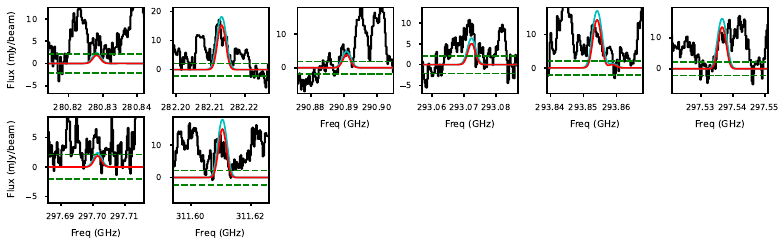}
\caption{Observations of the CH$_3$OH transitions expected to be optically thin at the continuum peak position (in black) compared to the best-fit model when fixing the parameters $s$, $T_{\rm ex}$, and FWHM in red. The $\varv_{\rm LSR}$ is free. In cyan, we show the model with the column density derived from CH$_3$$^{18}$OH, $N$ = 5\,$\times$\,10$^{19}$ cm$^{-2}$. The dashed green lines correspond to $\pm$1 rms (per channel).}
\label{fig_model_CH3OH}
\end{center}
\end{figure*}

\begin{figure*}[h!]
\begin{center}
\includegraphics[width=16cm]{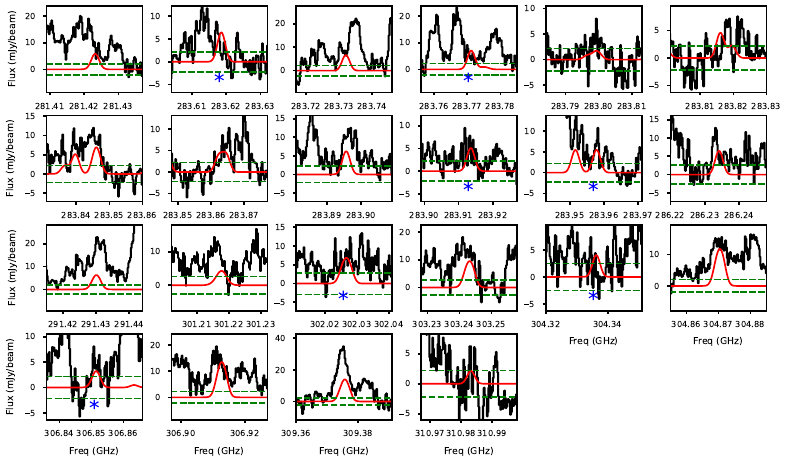}
\caption{Observations of the CH$_3$$^{17}$OH transitions at the continuum peak position including the ones partially blended with other molecules (in black) compared to the best-fit model in red using the same parameters $s$, $T_{\rm ex}$, FWHM, and $\varv_{\rm LSR}$ as for the other isotopologues. The dashed green lines correspond to $\pm$1 rms (per channel). The lines that are used in the $\chi^2$ minimization are indicated with a blue $*$ below the lines.}
\label{fig_model_CH317OH}
\end{center}
\end{figure*}

\begin{figure*}[h!]
\begin{center}
\includegraphics[width=16cm]{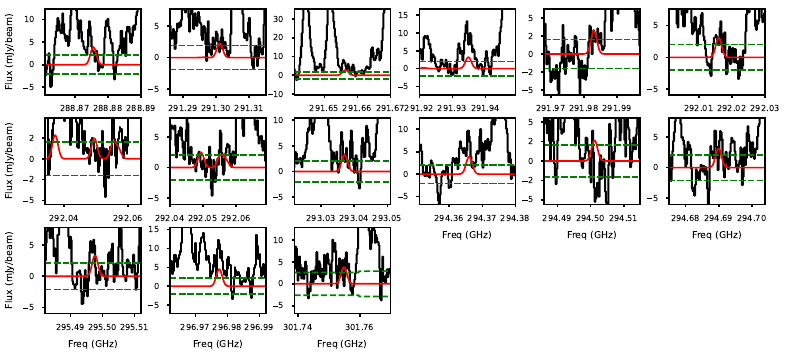}
\caption{Observations of the CD$_3$OD transitions at the redshifted position (in black) compared to the upper limit model (in red).}
\label{fig_model_CD3OD}
\end{center}
\end{figure*}

\end{appendix}

\end{document}